\documentclass[%
reprint,
superscriptaddress,
amsmath,
amssymb,
aps,
pre,
floatfix,
]{revtex4-2}

\usepackage{graphicx}
\usepackage{dcolumn}
\usepackage{bm}
\usepackage{hyperref}

\usepackage[utf8]{inputenc}
\usepackage[T1]{fontenc}
\usepackage{mathptmx}
\usepackage{gensymb} 

\usepackage{amsmath}
\usepackage{amssymb}
\usepackage{graphicx}
\usepackage{dcolumn}
\usepackage{bm}
\usepackage{multirow}
\usepackage{hyperref}
\usepackage{subfigure}
\usepackage{adjustbox}
\usepackage{tabularx}

\DeclareSymbolFont{epsilon}{OML}{cmm}{m}{it}

\begin{document}

\preprint{APS/123-QED}

\title{The effect of stereochemical constraints on the structural properties of folded proteins}

\author{Jack A. Logan}
\thanks{These authors contributed equally to this work.}
\affiliation{Department of Mechanical Engineering, Yale University, New Haven, Connecticut 06520, USA}
\author{Jacob Sumner}
\thanks{These authors contributed equally to this work.}
\affiliation{Graduate Program in Computational Biology and Bioinformatics, Yale University, New Haven, Connecticut, 06520, USA}
\affiliation{Integrated Graduate Program in Physical and Engineering Biology, Yale University, New Haven, Connecticut, 06520, USA}
\author{Alex T. Grigas}
\affiliation{Graduate Program in Computational Biology and Bioinformatics, Yale University, New Haven, Connecticut, 06520, USA}
\affiliation{Integrated Graduate Program in Physical and Engineering Biology, Yale University, New Haven, Connecticut, 06520, USA}
\author{Mark D. Shattuck}
\affiliation{Benjamin Levich Institute and Physics Department, The City College of New York, New York, New York 10031, USA}
\author{Corey S. O'Hern}
\affiliation{Department of Mechanical Engineering and Materials Science, Yale University, New Haven, Connecticut, 06520, USA}
\affiliation{Graduate Program in Computational Biology and Bioinformatics, Yale University, New Haven, Connecticut, 06520, USA}
\affiliation{Integrated Graduate Program in Physical and Engineering Biology, Yale University, New Haven, Connecticut, 06520, USA}
\affiliation{Department of Physics, Yale University, New Haven, Connecticut, 06520, USA}
\affiliation{Department of Applied Physics, Yale University, New Haven, Connecticut, 06520, USA}

\date{\today}

\begin{abstract}
Proteins are composed of chains of amino acids that fold into complex three-dimensional structures. Several key features, such as the radius of gyration, fraction of core amino acids $f_{\rm core}$, packing fraction $\langle \phi\rangle$ of core amino acids, and structure factor $S(q)$ define the structure of folded proteins. It is well-known that folded proteins are compact with a radius of gyration $R_g(N) \sim N^{\nu}$ that obeys power-law scaling with the number of amino acids $N$ and $\nu \sim 1/3$, $f_{\rm core} \approx 0.09$, and $\langle \phi \rangle \approx 0.55$. We also investigate the {\it internal} scaling of the radius of gyration $R_g(n)$ versus the chemical separation $n$ between amino acids for subchains of length $n$ and show that it does not obey simple power-law scaling with $\nu \sim 1/3$.  Instead, $R_g(n) \sim n^{\nu_{1,2}}$ with a larger exponent $\nu_1 > 1/3$ for small $n$ and smaller exponent $\nu_{2} < 1/3$ for large $n$. To develop a minimal model for proteins that recapitulates these defining structural features, we carry out collapse simulations for a series of coarse-grained models with increasing complexity. We show that a model, which coarse-grains amino acids into a single spherical backbone bead and several variable-sized side-chain beads and enforces bend- and dihedral-angle constraints for the backbone, recapitulates $R_g(n)$, $f_{\rm core}$, $\langle \phi \rangle$, and $S(q)$ for more than $2500$ x-ray crystal structures of proteins.
\end{abstract}


\maketitle

\section{Introduction \label{sec:introduction}}

Proteins are polypeptide chains containing tens to thousands of amino acids that carry out important cellular and extracellular functions. While breakthroughs in machine learning have improved our ability to predict the x-ray crystal structures of proteins from their amino acid sequences~\cite{abramson_accurate_2024, baek_accurate_2021, lin_evolutionary-scale_2023} and to design new protein sequences~\cite{dauparas_robust_2022}, modeling the physical and dynamic process of protein folding remains a challenge. In particular, experimental studies of protein folding have revealed intermediate kinetic traps, fold switching, mechanisms of misfolding and aggregation, allostery, and structural changes in response to mutations~\cite{anfinsen_principles_1973, porter_extant_2018, dobson_protein_1999, monod_nature_1965, fersht_folding_1992, gruebele2016globular}, all of which still require theoretical and computational modeling.

\begin{figure*}[htbp]
    \centering
    \adjustbox{valign=B}{\subfigure{\normalsize (a)}}
    \adjincludegraphics[valign=T,width=0.73\columnwidth]{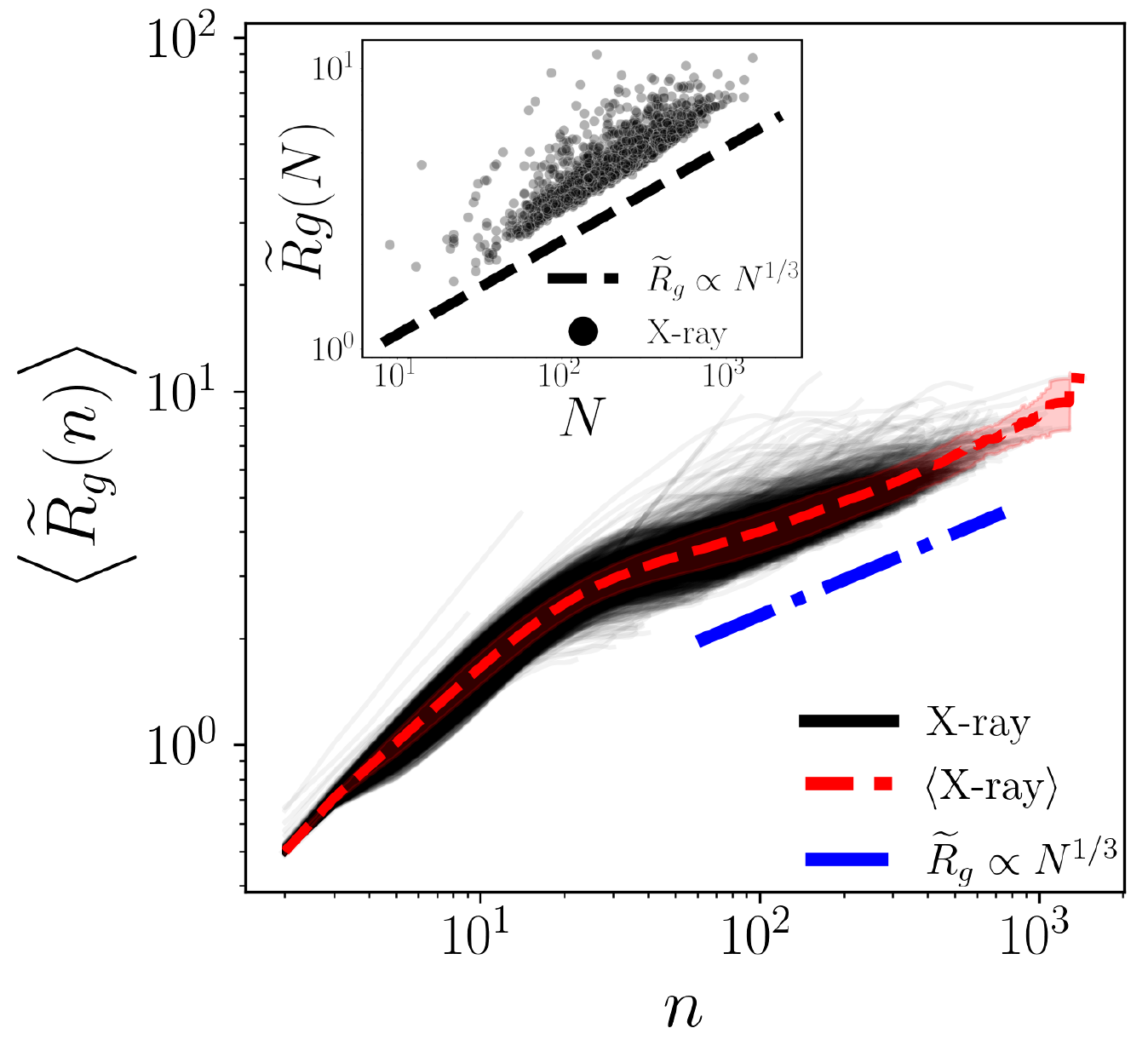}
    \hspace{-3mm}
    \adjustbox{valign=B}{\subfigure{\normalsize (b)}}
    \adjincludegraphics[valign=T,width=0.59\columnwidth]{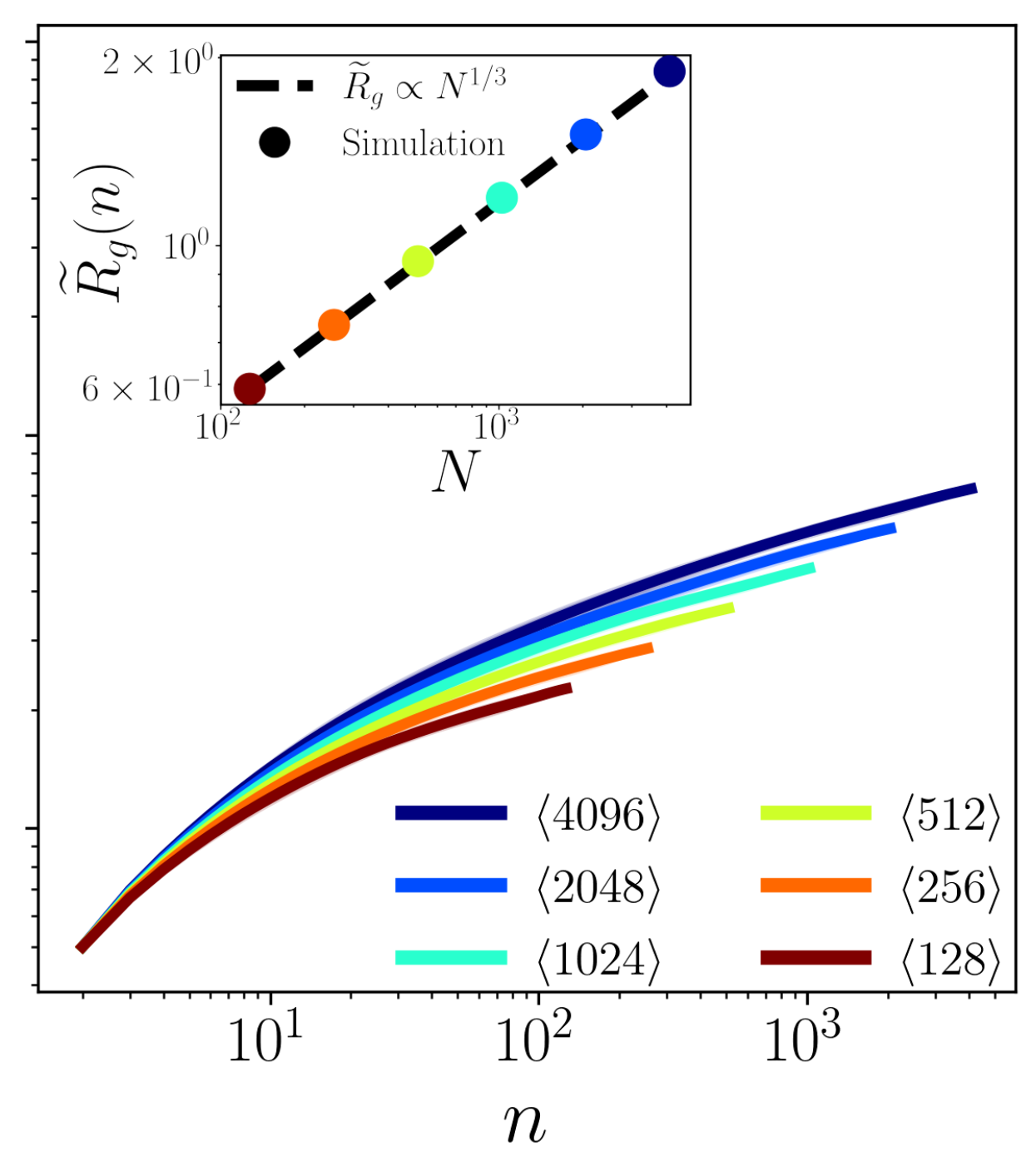}
    \hspace{-3mm}
    \adjustbox{valign=B}{\subfigure{\normalsize (c)}}
    \adjincludegraphics[valign=T,width=0.605\columnwidth]{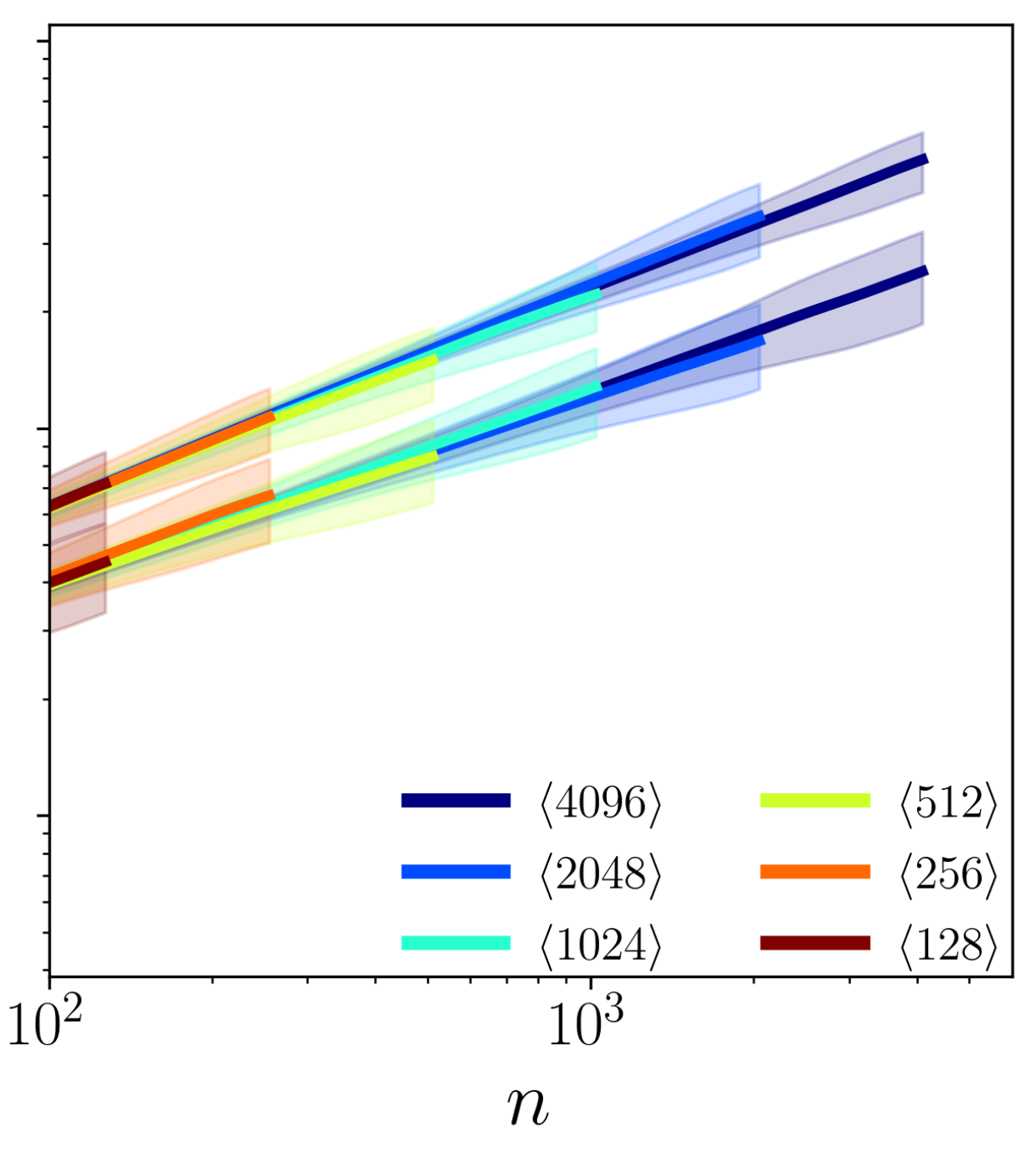}
    \caption{Average normalized radius of gyration $\langle \widetilde{R}_g(n)\rangle$ as a function of the subchain length $n$. (a) The anomalous scaling of $\langle \widetilde{R}_g(n)\rangle$ for $2531$ x-ray crystal structures of single-chain proteins with variable numbers of amino acids $N$ (thin black lines). The dashed red line gives the average over all proteins. The dot-dashed blue line has a slope of $1/3$. In the inset, we show $\langle \widetilde{R}_g(N)\rangle$ for the same x-ray crystal structures (filled black circles). The dashed black line has a slope of $1/3$. (b) For collapsed, excluded-volume bead-spring polymers as for folded proteins, $\langle \widetilde{R}_g(n)\rangle$ does not obey power-law scaling behavior with a {\it single} exponent. However, in the inset, we show that the endpoints obey $\widetilde{R}_g(N) \propto N^{1/3}$ for $N=128$ (black line) to $4096$ (violet line) spherical monomers. (c) $\langle \widetilde{R}_g(n)\rangle \propto n^{\nu}$ with $\nu \sim 0.59$ for excluded-volume random-walk polymers (upper curves) compared to $\nu \sim 0.50$ for ideal random-walk polymers (lower curves).}
    \label{fig:internal_vs_endpoint_scaling}
\end{figure*}

Globular proteins fold into complex three-dimensional conformations with compact interiors, or core regions, that determine their thermal stability~\cite{dill1990dominant}. Previous studies have shown that the fraction of amino acids in protein cores $f_{\rm core} \approx 0.09$, and the average packing fraction of core amino acids (without non-bonded atomic overlaps) $\langle \phi \rangle \approx 0.55$~\cite{treado2019void, Mei2020NMR, grigas_using_2020, grigas_core_2022, grigas_protein_2025}. The overall structure of folded proteins can be characterized by the structure factor $S({\vec q})=N^{-1} \sum_{k=1}^N \sum_{l=1}^{N} e^{i {\vec q} \cdot ({\vec r}_k-{\vec r}_l)}$ and radius of gyration of the protein backbone,
\begin{equation}
    R_g(N) = \sqrt{\frac{1}{N}\sum_{k=1}^N |\vec{\mathbf{r}}_k - \vec{\mathbf{r}}_{\mathrm{com}}|^2},
\end{equation}
where ${\vec q}$ is the wavevector, $\vec{\mathbf{r}}_k$ are the positions of the $N$ C$_{\alpha}$ atoms in the protein, and $\vec{\mathbf{r}}_{\mathrm{com}}$ is its center of mass. Both $S(q)$ and $R_g(N)$ have been employed as reaction coordinates for the folding process~\cite{yamamoto2021universal} and used to identify intrinsically disordered proteins (IDPs), which do not adopt a single compact structure, but contain both open and compact regions~\cite{smith2014calibrated}.

The radius of gyration for simple polymers follows power-law scaling relations, $R_g(N) \propto N^\nu$, where $\nu = 1$ for fully extended polymers, $0.5$ for random-walk polymers, and $1/3$ for collapsed polymers. Recent studies of x-ray crystal structures of globular proteins have shown that $R_g(N) \sim N^{\nu^*}$ with exponent $\nu^* \sim 0.33$-$0.4$~\cite{hong2009scaling}, similar to the behavior for collapsed polymers. (See the inset to Fig.~\ref{fig:internal_vs_endpoint_scaling} (a).) Deviations from the power-law scaling behavior with exponent $\nu^*$ are found for proteins with small ratios of hydrophobicity to electric charge~\cite{damaschun1992streptokinase,lattman1994small,kataoka1996x,durchschlag1996comparative,men2005structural,biehl2011exploring}. However, proteins with similar $N$, can possess strongly differing conformations. 

To gain additional insight into the internal structure of proteins, we can define $\langle R_g(n)\rangle$ as the average radius of gyration over all subchains of length $n \le N$,
\begin{equation}     
\label{eq:Rg(n)_definition}
\langle R_g(n) \rangle = \frac{1}{N-n} \sum\limits_{i=1}^{N-n}  R_{g}(i, i+n-1),
\end{equation}
where
\begin{equation}
R_g(i,j) = \left[ \frac{1}{j-i+1} \sum\limits_{k=i}^{j} \left( \vec{\mathbf{r}}_k - \langle \vec{\mathbf{r}}_k \rangle \right)^2 \right]^{1/2}
\end{equation}
and 
\begin{equation}
\langle \vec{\mathbf{r}}_k \rangle = \frac{1}{j-i+1} \sum\limits_{k=i}^{j} \vec{\mathbf{r}}_k.
\end{equation}
In Fig.~\ref{fig:internal_vs_endpoint_scaling} (a), we show that while the $R_g(N)$ scaling for folded proteins obeys $R_g(N) \propto N^{\nu^*}$ with $\nu^* \sim 0.33$-$0.4$, the internal scaling $R_g(n)$ is more complex. $R_g(n)$ possesses two characteristic power-law scaling regions: $R_g(n) \propto n^{\nu_{1,2}}$ with $\nu_1 \sim 0.7 > 1/3$ for small $n$ and $\nu_2 \sim 0.2 < 1/3$ for large $n$, which differs significantly from $R_g(n)$ for collapsed bead-spring polymers (Fig.~\ref{fig:internal_vs_endpoint_scaling} (b)), as well as excluded-volume and ideal random-walk polymers (Fig.~\ref{fig:internal_vs_endpoint_scaling} (c)). 

We seek to develop a minimal model for proteins that captures the key structural properties observed in high-resolution x-ray crystal structures of folded proteins. All-atom models have been used to fold proteins computationally~\cite{mccammon_dynamics_1977, karplus_protein_1979, brunger_crystallographic_1987, karplus_molecular_2002, lindorff-larsen_systematic_2012, duan_pathways_1998, zhou_free_2001, lindorff-larsen_how_2011, piana_protein_2012, voelz_molecular_2010}, yet they have only folded proteins with $N \lesssim 100$ and typically capture only $\mu$s to $m$s time scales~\cite{shaw_atomic-level_2010, freddolino_challenges_2010, noe_beating_2015}. Coarse-grained models can potentially be used to fold larger proteins by reducing the geometric complexity of the amino acids. Coarse-grained models for proteins range from one spherical bead per amino acid~\cite{levitt_computer_1975,berger_protein_1998,thirumalai_deciphering_1999,clementi_topological_2000} to one spherical bead for the backbone and one or multiple beads for the side chains~\cite{liwo_ab_2005,sterpone_opep_2014,kmiecik_one-dimensional_2017,souza_martini_2021,davtyan_awsem-md_2012,kar_transferring_2014}. Prior coarse-grained models for proteins are typically calibrated by matching the radius of gyration $R_g(N)$ to within $10\%$ of the x-ray crystal structure or achieving root-mean-square deviation (RMSD) of the $C_{\alpha}$ atoms ${\rm RMSD} \lesssim 3$~\AA{} from the x-ray crystal structure. However, matching only these two metrics to the x-ray crystal structures does not ensure that the model protein captures the core structure of the x-ray crystal structure. 

Thus, in this work, we investigate a range of coarse-grained models of proteins to determine the minimal model that recapitulates four key properties that define the structure of folded proteins: $\langle R_g(n)\rangle$, $\langle \phi \rangle$, $f_{\rm core}$, and $S(q)$. We focus on six coarse-grained protein models with increasing complexity: a collapsed excluded-volume bead-spring random-walk polymer model, the previous polymer model with effective bend- and dihedral-angle constraints, the previous polymer model with an additional side-chain spherical bead attached to each backbone spherical bead, the previous polymer model except the sizes of each side-chain spherical beads are selected to mimic the side chains of amino acids in the protein, the previous polymer model with the same side chain representations as those employed in Martini3~\cite{souza_martini_2021}, and the previous polymer model with the single side-chain spherical beads of leucine and valine replaced with multiple side chain beads. To simplify the polymer models, we do not include explicit attractive interactions between amino acids. Instead, to induce hydrophobic collapse of the coarse-grained protein models, we employ an external compressive central force with damped molecular dynamics (MD) simulations. Previous studies have shown that the structural properties of bead-spring polymers collapsed using attractive interactions are similar to those for purely repulsive bead-spring polymers compressed using a central force~\cite{grigas2024connecting}. In addition, static packings of purely repulsive, rigid, amino acid-shaped particles compressed to jamming onset (i.e. the maximum packing fraction that does not give rise to overlaps between amino acids) achieve a similar average packing fraction as that found in the cores of x-ray crystal structures of globular proteins~\cite{gaines2016random,Mei2020NMR}. 

Below, we describe the results for simulations of chain collapse for all six coarse-grained models for more than $2500$ individual proteins (with $N=100$–$1500$).  We show that models with sufficiently complex side‐chain representations accurately reproduce $\langle R_g(n)\rangle$, $\langle \phi \rangle$, $f_{\rm core}$, and $S(q)$ over the full data set of proteins. In future studies, the accurate coarse-grained models described here can potentially be used for folding proteins of unknown structure, docking protein monomers to determine protein-protein interactions, and other protein structure prediction applications.

This article is organized into three additional sections and four appendices. In Sec.~\ref{sec:methods}, we describe the six coarse-grained protein models and the simulation protocol for studying protein chain collapse. In Sec.~\ref{sec:discussion}, we describe the results for $R_g(n)$, $S(q)$, $\langle \phi \rangle$, and $f_{\rm core}$ for each coarse-grained protein model and compare the results to those for the x-ray crystal structures of proteins. In Sec.~\ref{sec:conclusion}, we emphasize the importance of developing coarse-grained protein models that can accurately capture the structure of protein cores in x-ray crystal structures. We also outline future coarse-grained simulations that can recapitulate protein folding dynamics with small root-mean-square deviations from x-ray crystal structures of proteins. In Appendix~\ref{app:a}, we describe the constraints that we used to obtain the dataset of $\sim 2500$ x-ray crystal structures of proteins from the Protein Data Bank. In Appendix~\ref{app:b}, we provide the procedure for generating the initial conformations for each coarse-grained protein model. In Appendix~\ref{app:c}, we describe the dihedral angle potential energy function for the coarse-grained protein models. Finally, in  Appendix~\ref{app:d}, we illustrate the method that we employ to identify the core residues and calculate $f_{\rm core}$ and $\langle \phi \rangle$ in both the x-ray crystal structures and coarse-grained protein models.

\section{Methods \label{sec:methods}}

\begin{figure}[htbp]
    \centering
    \mbox{}%
    \adjustbox{valign=B}{\subfigure{(a)}}
    \adjincludegraphics[valign=T,width=0.39\columnwidth]{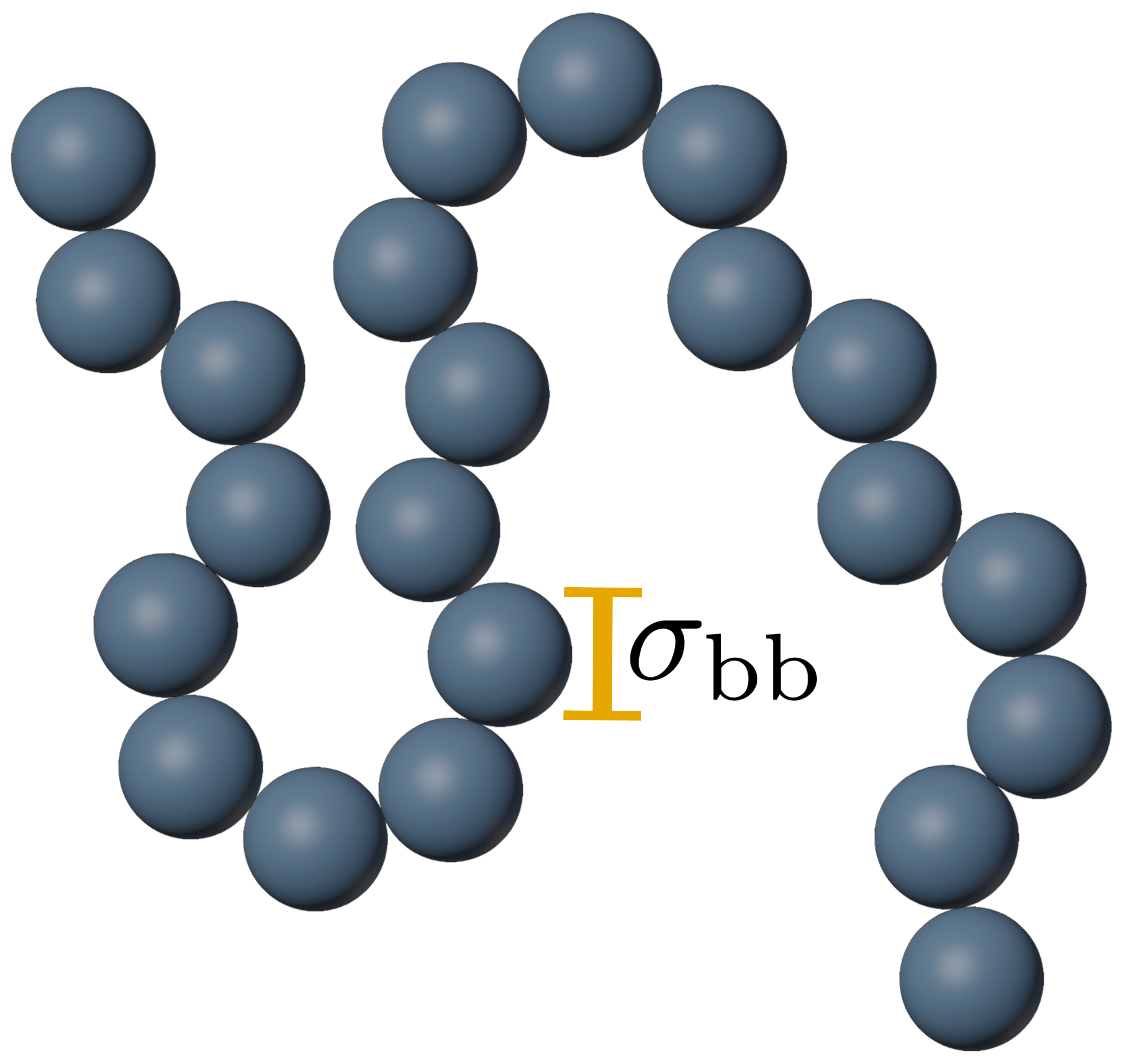}
    \hspace{5mm}
    \adjustbox{valign=B}{\subfigure{(b)}}
    \adjincludegraphics[valign=T,width=0.37\columnwidth]{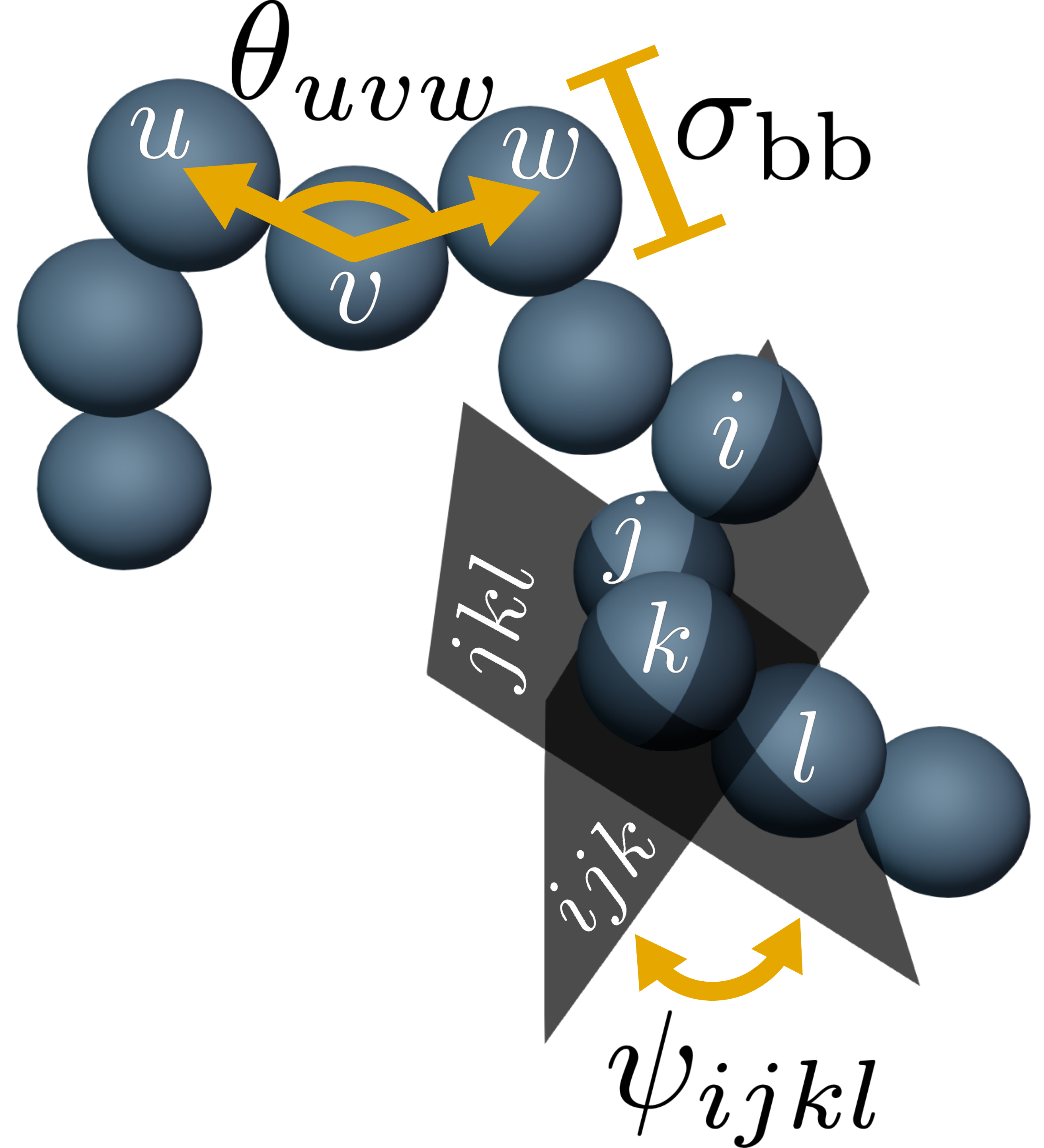}
    \par\medskip
    \adjustbox{valign=B}{\subfigure{(c)}}
    \adjincludegraphics[valign=T,width=0.39\columnwidth]{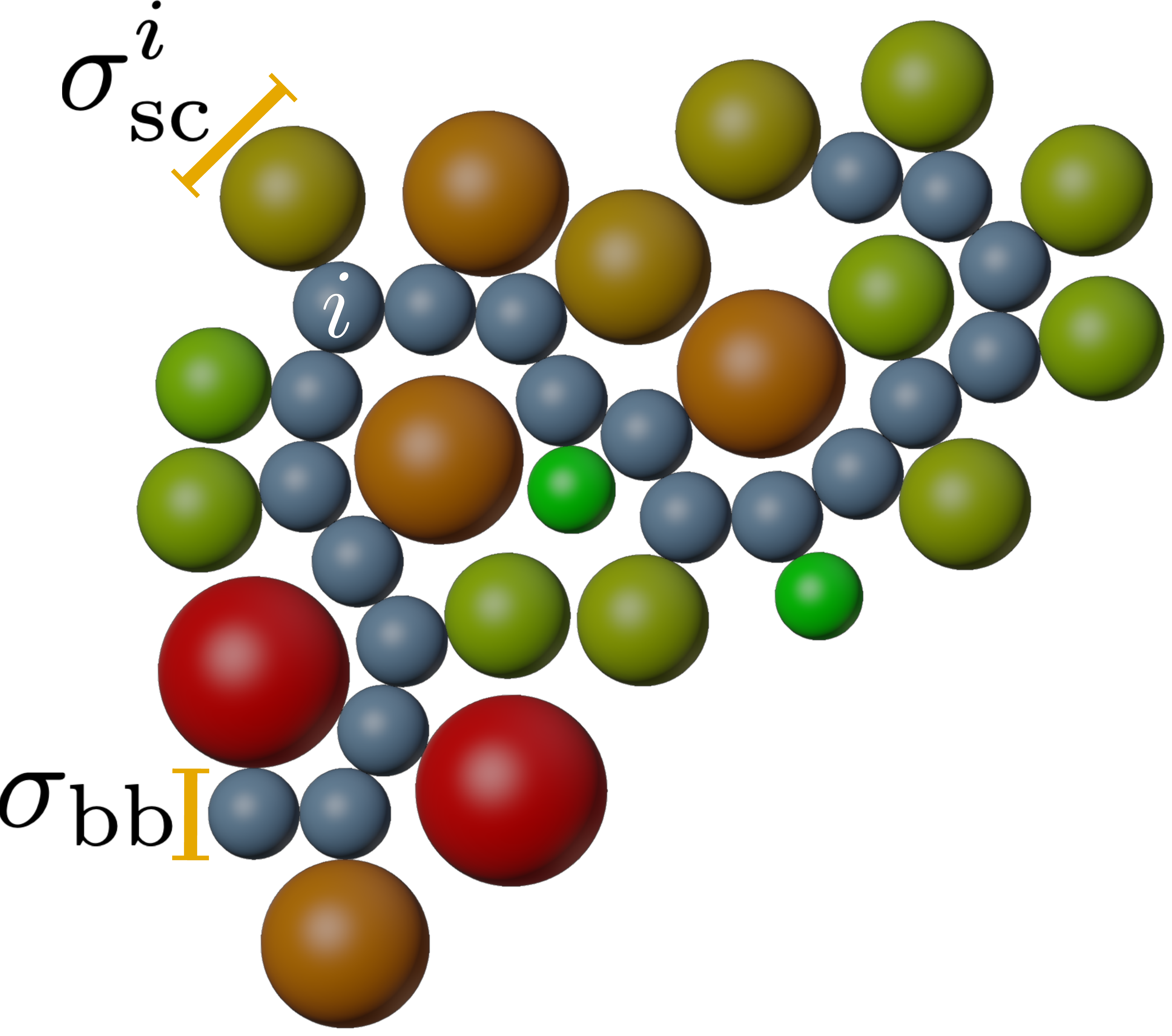}
    \hspace{5mm}
    \adjustbox{valign=B}{\subfigure{(d)}}
    \adjincludegraphics[valign=T,width=0.39\columnwidth]{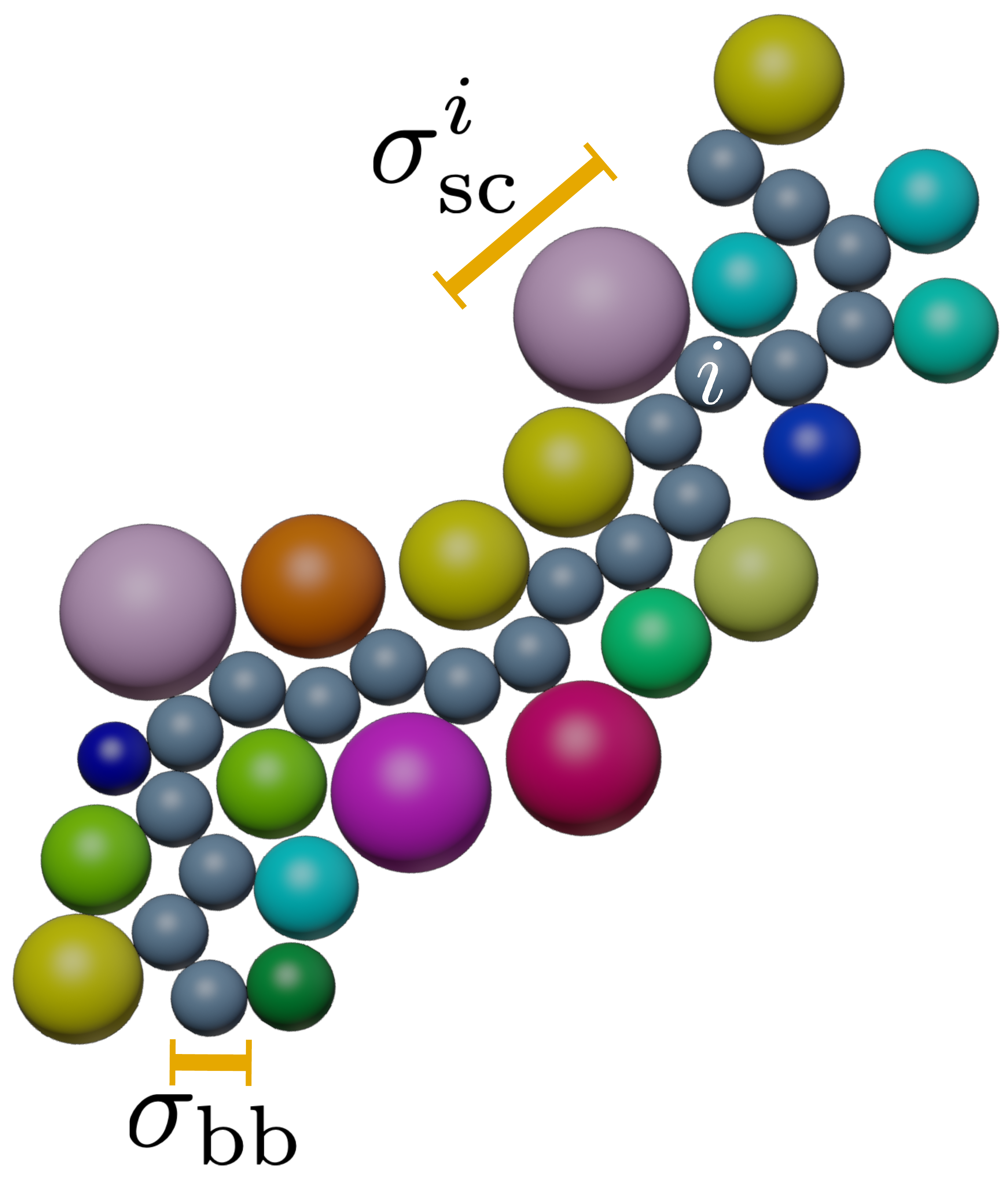}
    \par\medskip
    \adjustbox{valign=B}{\subfigure{(e)}}
    \adjincludegraphics[valign=T,width=0.41\columnwidth]{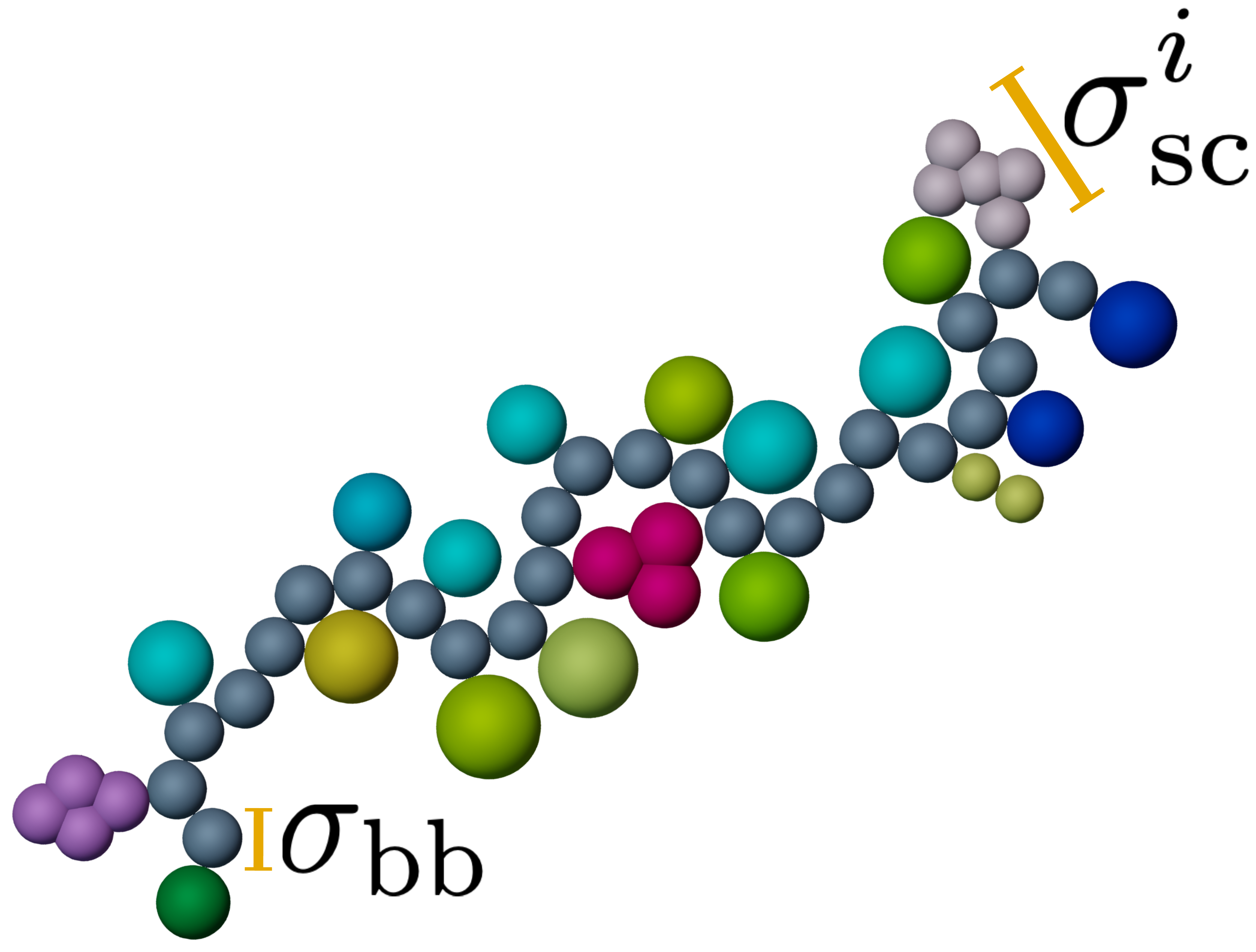}
    \hspace{5mm}
    \adjustbox{valign=B}{\subfigure{(f)}}
    \adjincludegraphics[valign=T,width=0.39\columnwidth]{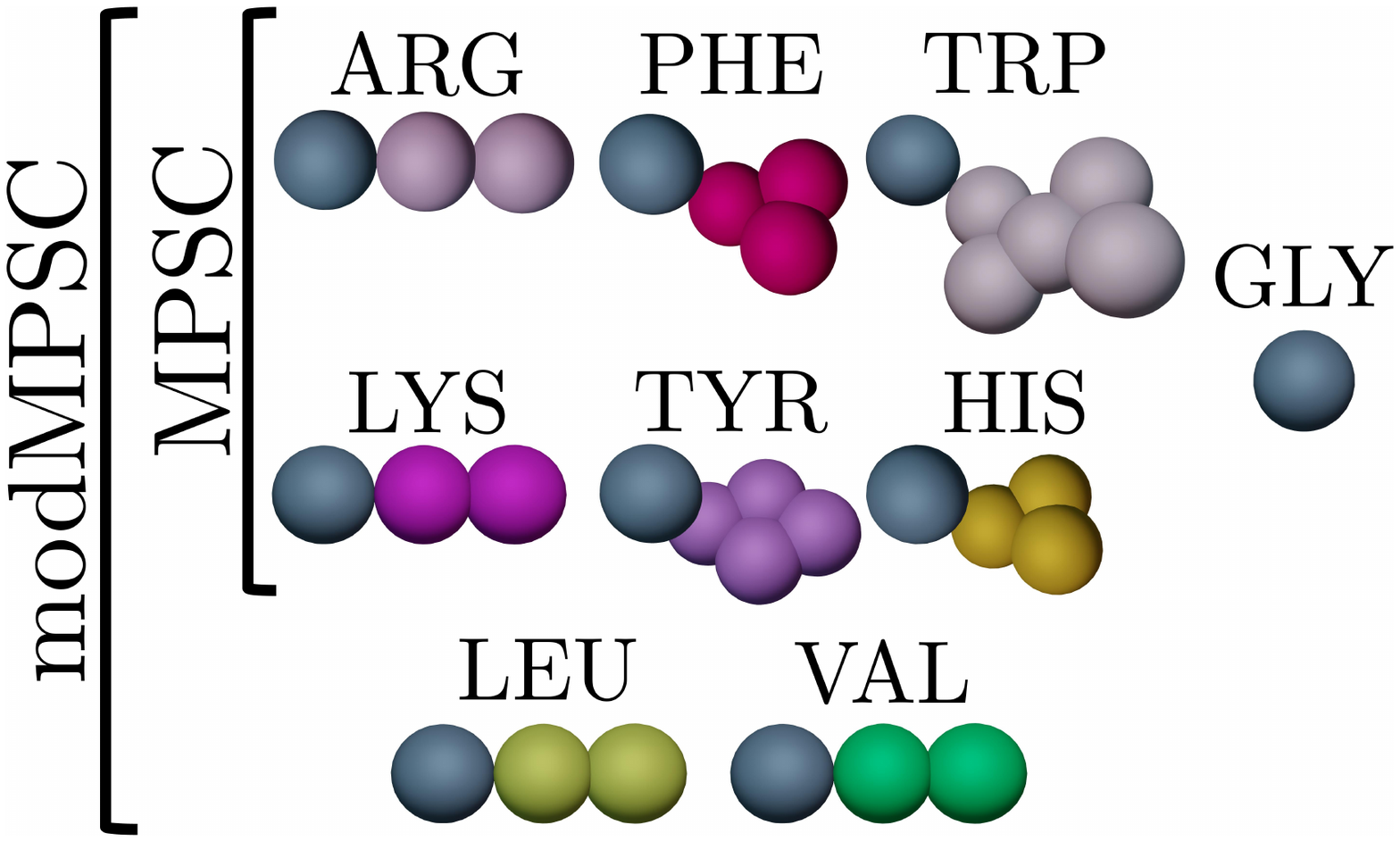}
    \caption{(a)-(e) Snapshots of the six coarse-grained models of proteins, shown as 2D projections. When moving from (a)-(e), the successive models include all features of the previous models. $\sigma_{\mathrm{bb}}$ indicates the diameter of the spherical bead that represents the backbone of each amino acid. (a) A collapsed freely-jointed excluded-volume random walk (CRW) polymer chain with inter-amino acid separation $\sigma_{\rm bb}$. (b) For the bend- and dihedral-angle potential (BADA) polymer model, the effective bend angles $\theta_{uvw}$ between three consecutive amino acids are constrained to values determined by x-ray crystal structures of proteins by a harmonic potential $U_{\rm bend}$, and the effective dihedral angles $\psi_{ijkl}$ between four consecutive amino acids are constrained to values determined by x-ray crystal structures of proteins by the dihedral angle potential $U_{\rm dh}$. (c) The freely jointed side-chain polymer model (FJSC) includes an additional spherical bead with diameter $\widetilde{\sigma}_{\mathrm{sc}}^{i}$ (colored by size) chosen randomly from a distribution of amino acid side chain diameters from x-ray crystal structures of proteins that are freely-jointed to each backbone monomer $i$. (d) For the ``in-sequence'' FJSC (In Seq) polymer model, the diameter of the side chain bead (colored by amino acid) is determined by the amino acid sequence that it is modeling. (e) The multi-particle side chain (MPSC) and modified MPSC (modMPSC) models use the geometry of the Martini3 side chains for seven types of amino acids. The modMPSC model differs from the MPSC model in using two spherical beads with a bend angle of $180^\circ$ for the side chains of LEU and VAL. (f) A summary of the amino acid side chain representations for the MPSC and modMPSC models. All other amino acids in these models have a single bead representation for the side chain, as for the In Seq model. The examples in (d)-(e) are sections of the protein, PDBID: 3ZZO.}
    \label{fig:simulation_models}
\end{figure}

In Fig.~\ref{fig:simulation_models}, we illustrate six coarse-grained models of proteins~\cite{taketomi_studies_1975,bahar_direct_1997,derreumaux_polypeptide_1999,kolinski_protein_2004,kar_primo_2013,davtyan_awsem-md_2012}. Each model has a connected backbone including one spherical bead per amino acid backbone with the same average separation between successive C$_{\alpha}$ atoms in proteins, i.e. $\sigma_{bb}\approx 3.8$ \AA. In order of increasing complexity, the models are: 1) a collapsed freely-jointed excluded-volume random-walk (CRW) polymer model, 2) the previous polymer model with constrained effective bend and dihedral angles (BADA) among the backbone spherical beads, 3) the previous polymer model with an additional spherical bead with randomly chosen diameter that is freely-jointed to each backbone monomer to represent the side chain for each amino acid (FJSC), 4) an ``in-sequence'' freely-jointed side chain polymer model (In Seq), where the diameter of the side chain bead mimics the size of the side chain of the protein's amino acid sequence, 5) a multi-particle side chain (MPSC) model similar to that for Martini3~\cite{souza_martini_2021}, where six of the amino acids contain more than one spherical side-chain bead and glycine does not have a side-chain bead, and 6) a modified MPSC model (modMPSC), where leucine and valine have two spherical side-chain beads. For each model, we perform more than $2500$ independent simulations, one for each protein in a dataset of high-resolution x-ray crystal structures of single-chain proteins~\cite{wang2003pisces}. See Appendix \ref{app:a} for more information about how we constructed the dataset used in this study.

In Fig.~\ref{fig:simulation_models} (a), we illustrate the CRW polymer model, where each of the $N$ spherical beads represents an amino acid with diameter $\sigma_{\mathrm{bb}}$. Neighboring amino acids $i$ and $j=i+1$ are connected using the harmonic bond length potential,
\begin{equation}
    U_{\mathrm{bond}}(r_{ij}) = \frac{U_{\mathrm{bb}}}{2} \left(1 - \frac{r_{ij}}{\sigma_{ij}} \right)^2,
\label{eq:NN_backbone_PE}
\end{equation}
where $r_{ij}$ is the separation between amino acids $i$ and $j$, $U_{\rm bb}$ is the strength of the bond length potential, and $\sigma_{ij}$ is the sum of the radii of the bonded monomers $i$ and $j$, $\sigma_{ij} = (\sigma_i + \sigma_j)/2$. Non-bonded amino acids interact via the purely repulsive linear spring potential,
\begin{equation}
    U_{\mathrm{rep}}(r_{ij}) = \frac{\epsilon_{\mathrm{rep}}}{2} \left(1 - \frac{r_{ij}}{\sigma_{ij}} \right)^2 \Theta \left(1 - \frac{r_{ij}}{\sigma_{ij}} \right),
\label{eq:non_backbone_rep_PE}
\end{equation}
where $\Theta(\cdot)$ is the Heaviside step function and $\epsilon_{\rm rep}$ is the strength of the non-bonded repulisve interactions between amino acids. Physical quantities will be made dimensionless using the energy scale $\epsilon_{\rm rep}$, the mass $m$ of an amino acid backbone bead, and the lengthscale $\sigma_{\rm bb}$.  Throughout this work a tilde over a given symbol is used to denote dimensionless quantities, e.g. $\widetilde{U}_{\mathrm{bb}} = U_{\mathrm{bb}}/\epsilon_{\mathrm{rep}}$. All dimensionless simulation parameters are defined in Appendices \ref{app:b} and \ref{app:c}.

In Fig.~\ref{fig:simulation_models} (b), we show that the BADA polymer model also includes constraints on the bend and dihedral angles between amino acids.  The bend angles $\theta_{ijk}$ between three sequential amino acids $i$, $j=i+1$, and $k=i+2$ are constrained by 
\begin{equation}
    U_{\mathrm{bend}}(\theta_{ijk}) = \frac{U_{\mathrm{ba}}}{2} \left(1 - \frac{\theta_{ijk}}{\theta_{ijk}^0} \right)^2,
\label{eq:bond_angles_PE}
\end{equation}
where the average bend angle $\theta^0_{ijk}$ is obtained from the x-ray crystal structure dataset. The 
dihedral-angle potential energy constrains the angle $\psi_{ijkl}$ between planes formed by the three beads $i$, $j$, and $k$ and three beads $j$, $k$, and $l$ among the four consecutive backbone beads $i$, $j$, $k$, and $l$:  
\begin{multline}
    U_{\mathrm{dh}}(\psi_{ijkl}) = U_{\mathrm{da}} \sum\limits_{\langle ijkl\rangle} \sum\limits_{s=1}^{4} \Bigl[ A_s \cos\left( s \, \psi_{ijkl}\right) \\ + B_s \sin\left( s \, \psi_{ijkl}\right)  \Bigr],
\label{eq:dihedral_PE}
\end{multline}
where $U_{\rm da}$ is the strength of the dihedral-angle potential and the dimensionless coefficients $A_s$ and $B_s$ are determined by the x-ray crystal structure dataset. (See Appendix \ref{app:c}.)

In Fig.~\ref{fig:angle_constraints} (a), we show the distribution ${\cal P}(\theta_{ijk})$ of bend angles between each set of three successive C$_{\alpha}$ atoms from the x-ray crystal structure dataset. The distribution has a strong peak around $\theta_{ijk} \approx 90^\circ$ and secondary peak near $120^\circ$. For each coarse-grained model that we simulate, we sample the bend angles randomly from ${\cal P}(\theta_{ijk})$, and then they are constrained using $U_{\rm bend}$ in Eq.~\ref{eq:bond_angles_PE}.  The dihedral-angle potential energy $U_{\mathrm{dh}}(\psi_{ijkl})$~\cite{smith2014calibrated,chen2006comparison}, as shown in Fig~\ref{fig:angle_constraints} (b), has a global minimum at $\psi_{ijkl} = \pm 180^\circ$, a peak near $~60^\circ$, and a plateau extending over the range $0^\circ ~\leq~\psi_{ijkl}~\leq~120^\circ$. Calculating the Boltzmann weight for $U_{\rm dh}$ yields ${\cal P}(\psi_{ijkl})$ for the x-ray crystal structure dataset. The key features in ${\cal P}(\theta_{ijk})$ and ${\cal P}(\psi_{ijkl})$ are attributed to protein secondary structure. The peak around $\theta_{ijk} \approx 90^\circ$ in ${\cal P}(\theta_{ijk})$ and the plateau in ${\cal P}(\psi_{ijkl})$ originate from $\alpha$-helical structures. The secondary peak near $\theta_{ijk} \approx 120^\circ$ and low-energy tails at $\psi_{ijkl} = \pm 180^\circ$ stem from $\beta$-sheet structures. Note that $\alpha$-helices are not favored by the coarse-grained dihedral-angle potential energy $U_{\rm dh}$.

\begin{figure}[htbp]
    \centering
    \mbox{}%
    \adjustbox{valign=B}{\subfigure{(a)}}
    \adjincludegraphics[valign=T,width=0.9\linewidth]{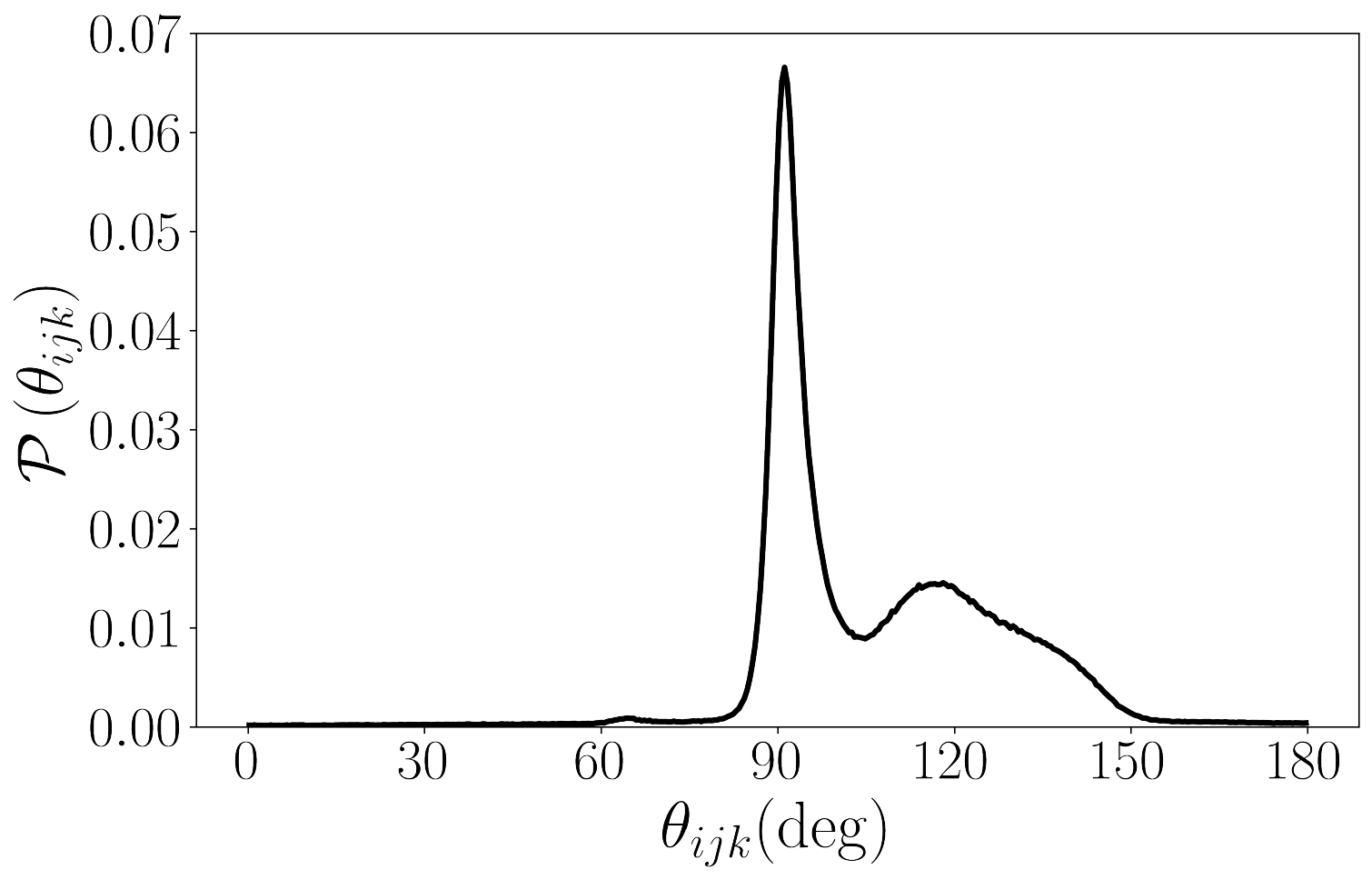}
    \par\medskip
    \adjustbox{valign=B}{\subfigure{(b)}}
    \adjincludegraphics[valign=T,width=0.9\linewidth]{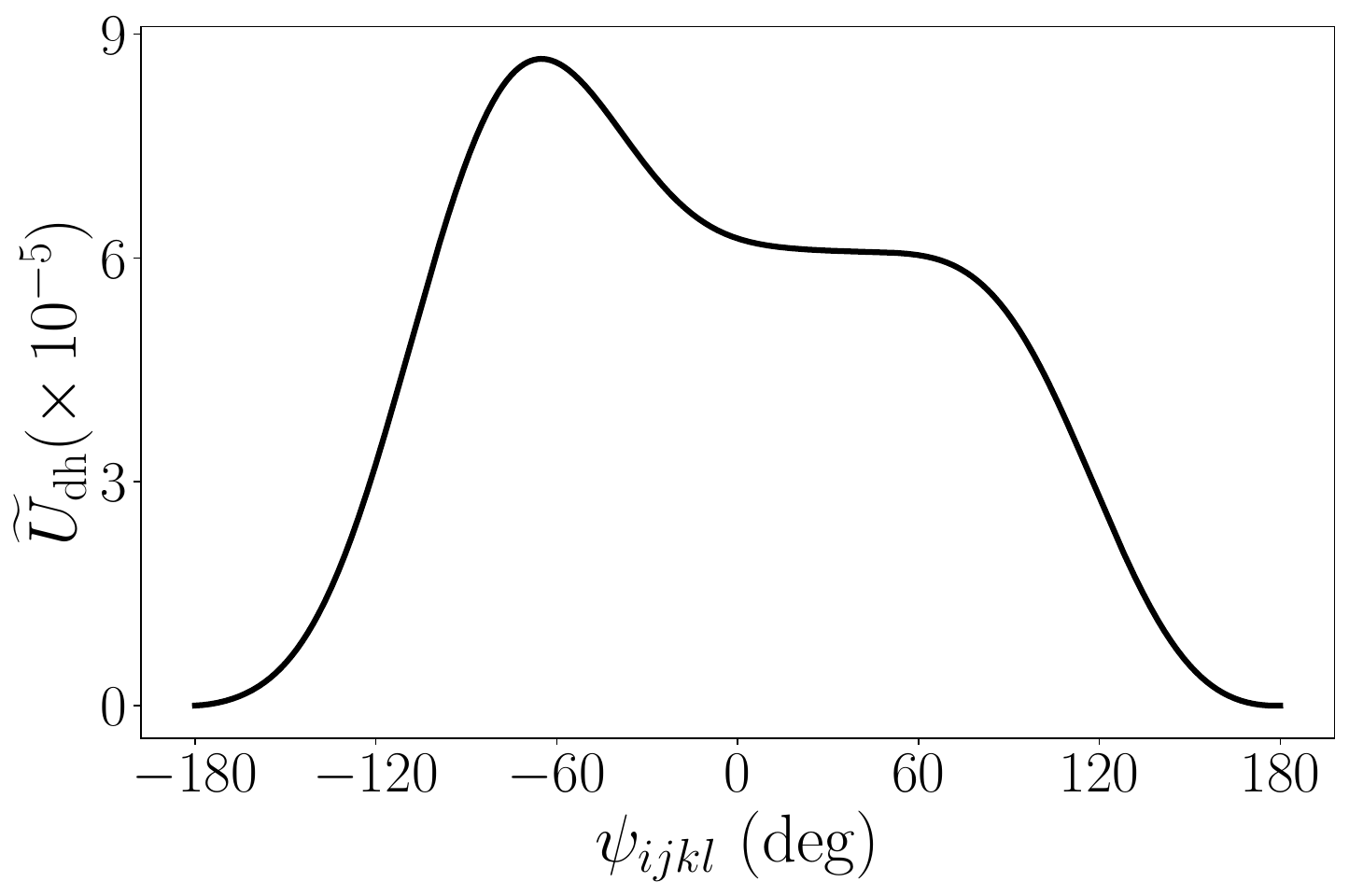}
    \caption{(a) Distribution ${\cal P}(\theta_{ijk})$ of the effective bend angles between three consecutive $C_{\alpha}$ atoms from the dataset of x-ray crystal structures of proteins. (b) The dimensionless dihedral-angle potential energy ${\widetilde U}_{\rm dh}(\psi_{ijkl})$ that yields the distribution ${\cal P}(\psi_{ijkl})$ of effective dihedral angles $\psi_{ijkl}$ between four consecutive C$_{\alpha}$ atoms observed in the x-ray crystal structure dataset when Boltzmann-weighting~\cite{smith2014calibrated}.}
    \label{fig:angle_constraints}
\end{figure}

\begin{figure}[htbp]
    \centering
    \mbox{}%
    \adjincludegraphics[valign=T,width=1\linewidth]{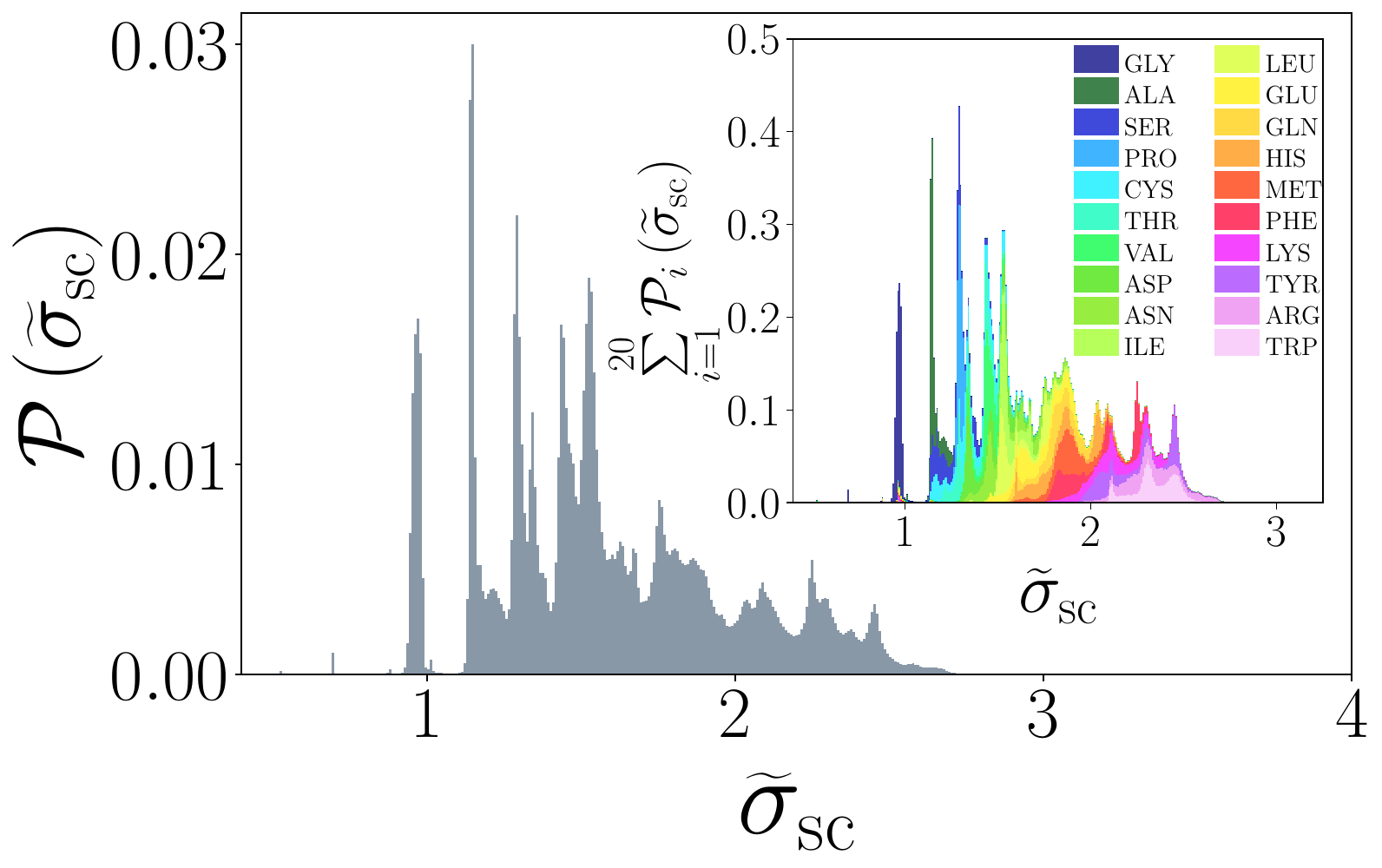}
    \caption{Distribution ${\cal P}(\widetilde{\sigma}_{\rm sc})$ of the effective side chain diameters (normalized by $\sigma_{\mathrm{bb}}$) binned over all amino acid types. The inset shows the sum of the distributions ${\cal P}_i({\widetilde \sigma}_{\rm sc})$ for each amino acid type $i$ indicated by different colors.}
    \label{fig:SC_diams}
\end{figure}

The coarse-grained protein models FJSC, In Seq, MPSC, and modMPSC in Fig.~\ref{fig:simulation_models} (c)-(f) incorporate side chain degrees of freedom, by freely-joining a spherical bead to each backbone bead (using Eq.~\ref{eq:NN_backbone_PE}). To approximate the effective diameter of each side chain, we calculate the maximum distance between all pairs of atoms in a side chain and add the average of the radii of the two atoms that are the farthest apart. The selected atomic radii have been used previously to calculate the average packing fraction of amino acids in protein cores~\cite{gaines2016random,gaines2017packing,treado2019void} and are provided in Appendix \ref{app:d}. Amino acid side chains can take on many conformations, so each amino acid possesses a distribution of effective side chain diameters. These distributions can either be calculated independently for each amino acid type or binned together to obtain an overall distribution of side chain diameters as shown in Fig.~\ref{fig:SC_diams}. For the FJSC polymer model in Fig.~\ref{fig:simulation_models} (c), the diameter $\sigma_{i,\mathrm{sc}}$ of the side chain bead bonded to backbone bead $i$ is chosen randomly from the overall distribution of effective amino acid side chain diameters ${\cal P}(\sigma_{\rm sc})$ in the main panel of Fig.~\ref{fig:SC_diams}. In contrast, for the In Seq polymer model in Fig.~\ref{fig:simulation_models} (d), we select the diameter of each side chain bead according to the amino acid sequence of each protein in the x-ray crystal structure dataset. In particular, the diameters of the side chain beads are randomly sampled from the individual amino acid side chain diameter distributions ${\cal P}_i(\sigma_{\rm sc})$ illustrated in the inset of Fig.~\ref{fig:SC_diams}, where ${\cal P}(\widetilde{\sigma}_{\mathrm{sc}}) = A \sum_{i=1}^{20} {\cal P}_i(\widetilde{\sigma}_{\mathrm{sc}}) / A_i$, $A$ is the normalization constant determined by $\int {\cal P}({\widetilde \sigma}_{\rm sc}) d{\widetilde \sigma}_{\rm sc}=1$, $A_i = 1 / (\Delta \widetilde{\sigma}_{\mathrm{sc}} N^i_c)$ is the normalization constant for the diameter distribution of amino acid $i$ with $N^i_c$ total counts and bin width $\Delta \widetilde{\sigma}_{\mathrm{sc}}$.

In Fig.~\ref{fig:simulation_models} (e), we show the MPSC model, which includes a single backbone spherical bead and side chains made up of $0$-$5$ spherical beads. The geometrical representations of the side chains are similar to those used in Martini3.  Glycine is now only represented by a backbone spherical bead. Each amino acid with a single side-chain sphere is unchanged from the In Seq model. Six amino acids in the MPSC model are represented by multiple side-chain spherical beads: arginine, phenylalanine, tryptophan, lysine, tyrosine, and histidine. The maximum dimension of the side chains with multiple spherical beads is the same as the diameter of the single side-chain bead representation in the In Seq model. To achieve this, the multiple side-chain spherical beads are the same size and rescaled so that the sum of the diameters matches the single side-chain bead diameter in the In Seq model. The modMPSC model is similar to the MPSC model, except the side chains for leucine and valine are represented by two spherical beads that form a $180^\circ$ bend angle with the backbone bead. In Fig.~\ref{fig:simulation_models} (f), we summarize the side-chain representations for the MPSC and modMPSC models.

When generating the initial coarse-grained protein conformations, the total potential energy contributions, $
U_{\text{rep}}^{\text{tot}} = \sum_{\langle i,j \rangle} U_{\text{rep}}(r_{ij}) \approx 0$ and $U_{\text{bond}}^{\text{tot}} = \sum_{\langle i,j \rangle} U_{\text{bond}}(r_{ij}) \approx 0$ for all models, and $U_{\text{bend}}^{\text{tot}} = \sum_{\langle i,j,k \rangle} U_{\text{bend}}(\theta_{ijk}) \approx 0$ for the BADA, FJSC, and In Seq models. We employ damped molecular dynamics (MD) simulations with an additional attractive central force on each bead to generate a collapsed conformation for each model and target protein. We employ a dimensionless damping parameter $\widetilde{\gamma} = 0.1$ in the overdamped limit, and run the damped MD simulations until the maximum magnitude of the net force on any bead $i$ satisfies $\text{max}_i {\widetilde F}_i < \widetilde{F}_{\text{tol}}$, where $F_i=|\vec{F}_i| = |{\vec \nabla}_{{\vec r}_i} U|$, $U$ is the total potential energy for a given model, and $\widetilde{F}_{\text{tol}} = 5\times10^{-13}$. We include an extra factor of the ratio of the bead diameter $\sigma_i$ to the maximum bead diameter $\sigma_{\mathrm{max}}$ raised to a power in the expression for the central force to ensure that the coarse-grained models do not form clusters of similar-sized beads during collapse when the beads are polydisperse~\cite{grigas2024connecting}:
\begin{equation}
    \vec{\mathbf{F}}_{\mathrm{cent}} = -F_{\mathrm{cent}} \left(\frac{\sigma_i}{\sigma_{\mathrm{max}}}\right)^{9/4}  \hat{\mathbf{r}}_i.
    \label{eq:central_force}
\end{equation}
The strength of the central force $\widetilde{F}_{\mathrm{cent}} = 10^{-4}$ compared to the constraint forces is such that the stereochemical constraints remain satisfied during collapse, e.g., the bend and dihedral angle distributions ${\cal P}(\theta_{ijk})$ and ${\cal P}(\psi_{ijkl})$ are nearly identical in the collapsed and initial states, and the results do not depend on $\widetilde{F}_{\rm cent}$.  We then calculate $\langle R_g(n)\rangle$, $\langle \phi \rangle$, $f_{\rm core}$, and $S(q)$ in the collapsed conformations for each coarse-grained protein model and protein target. 

\section{Results \label{sec:discussion}}

\begin{figure}[htbp]
    \centering
    \mbox{}%
    \adjustbox{valign=B}{\subfigure{(a)}}
    \adjincludegraphics[valign=T,width=0.9\linewidth]{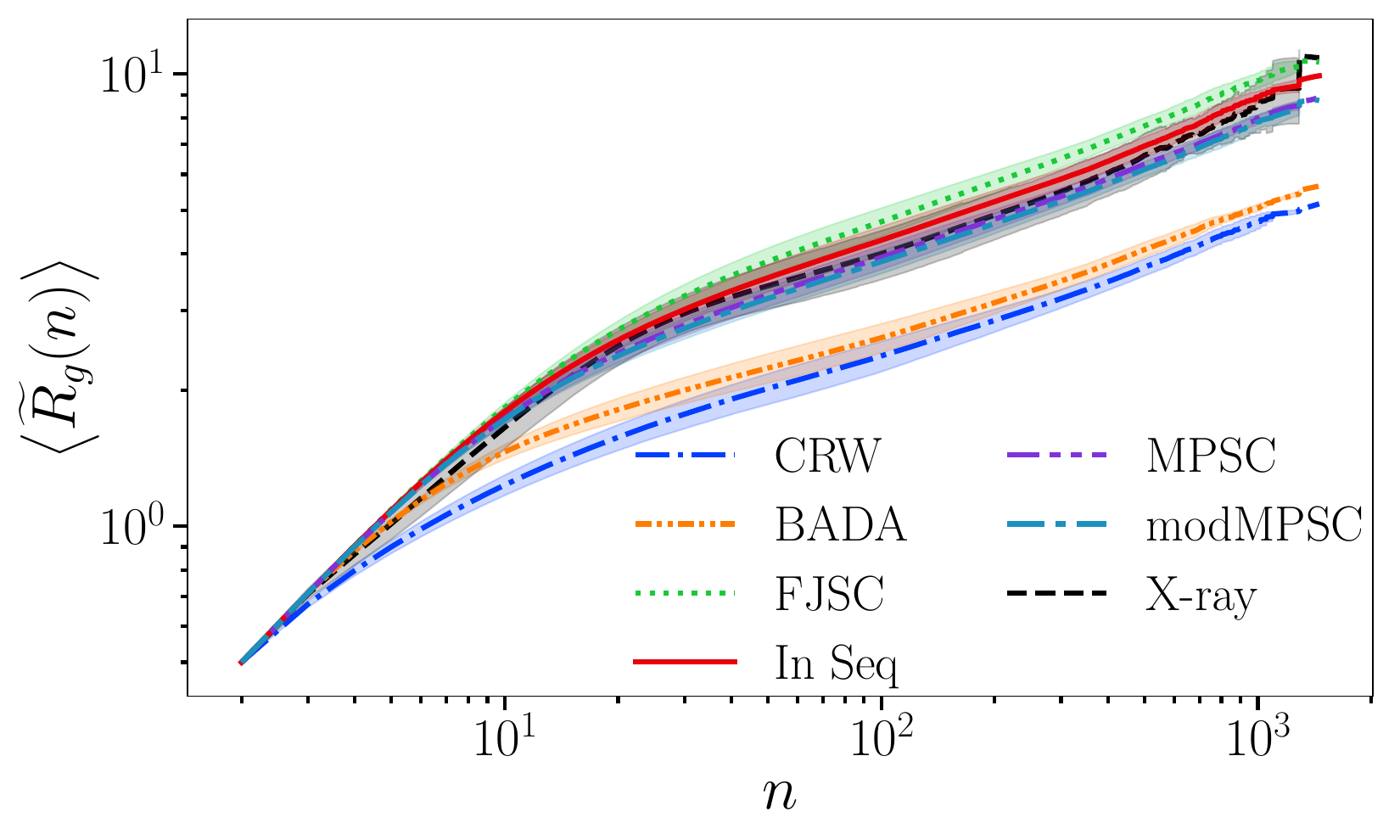}
    \par\medskip
    \adjustbox{valign=B}{\subfigure{(b)}}
    \adjincludegraphics[valign=T,width=0.9\linewidth]{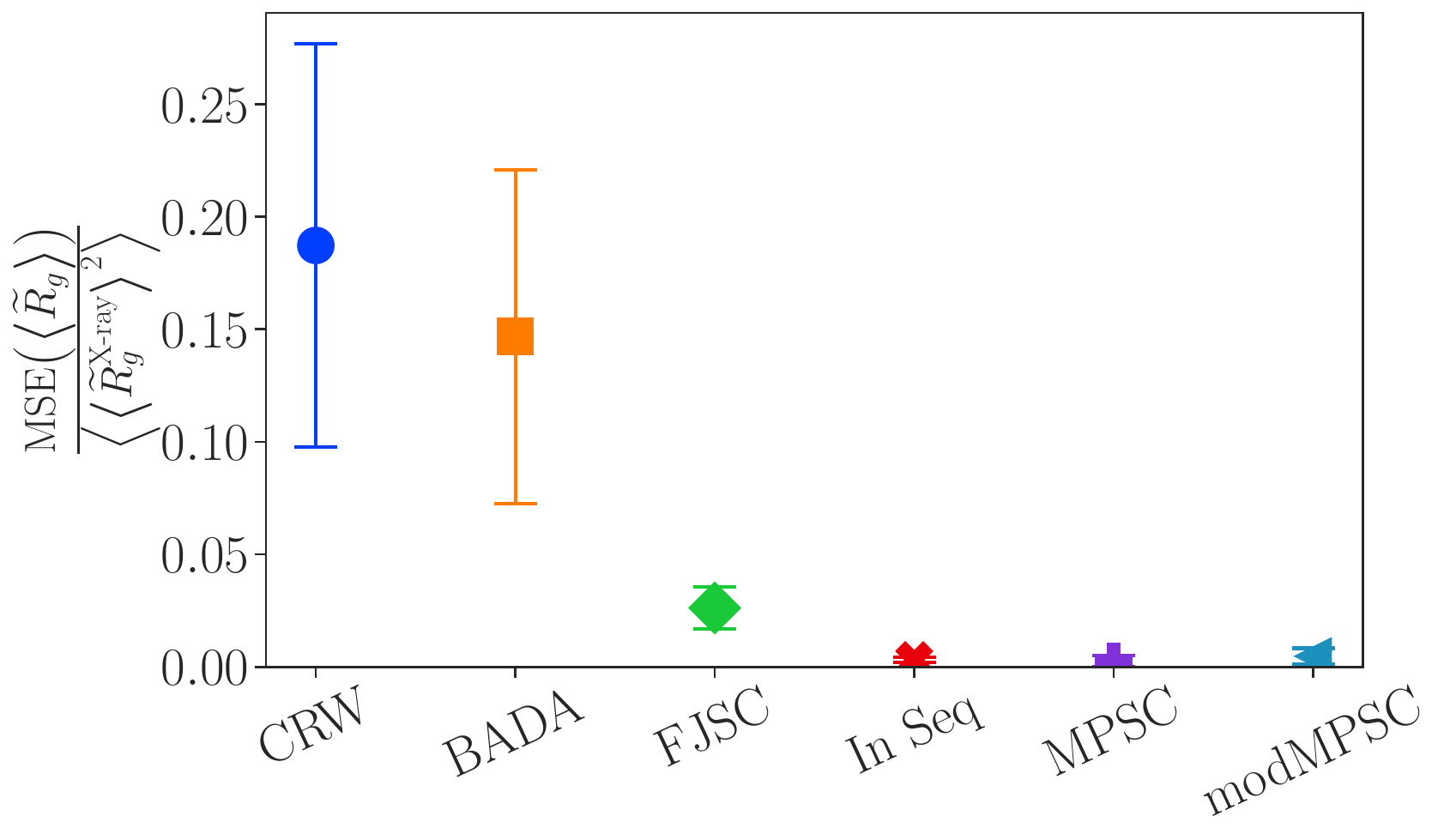}
    \par\medskip
    \adjustbox{valign=B}{\subfigure{(c)}}
    \adjincludegraphics[valign=T,width=0.9\linewidth]{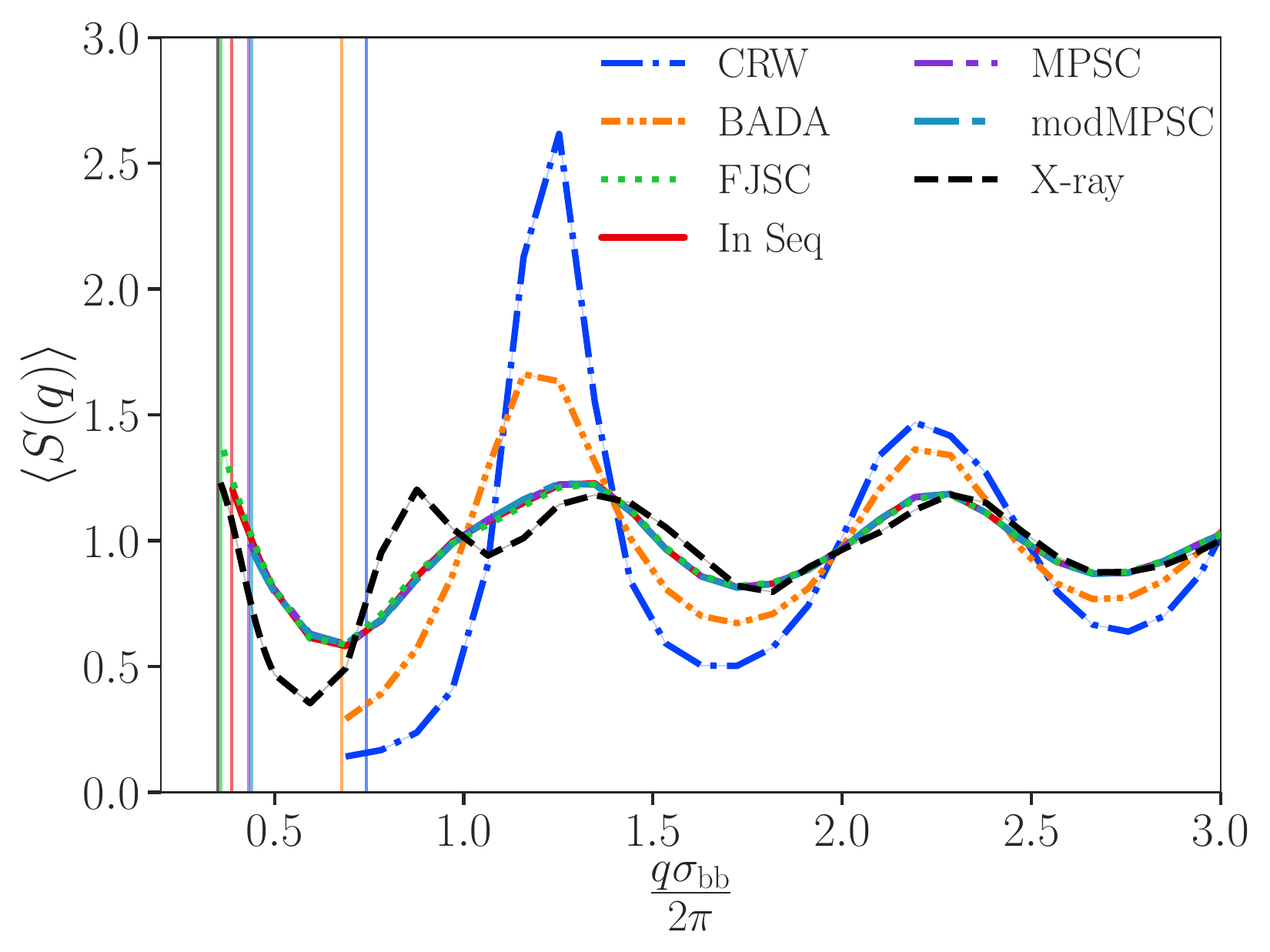}
    \caption{(a) The average radius of gyration $\langle \widetilde{R}_g(n) \rangle$ plotted versus subchain length $n$ for the x-ray crystal structures (black dashed line) and coarse-grained protein models with corresponding colors and line styles in the legend. The shading indicates the standard deviation about $\langle \widetilde{R}_g \rangle$ for each dataset. (b) Normalized mean-squared error in Eq.~\ref{mse} between $\langle \widetilde{R}_g(n) \rangle$ for each model and the average over the x-ray crystal structures. (c) The average structure factor $\langle S(q) \rangle$ plotted versus the wavenumber $q$ scaled by the diameter of the coarse-grained backbone size $\sigma_{bb}$. The vertical lines indicate the wavenumbers $q=2\pi/\mathrm{max}\{\langle \widetilde{R}_g(N) \rangle \}$ for each polymer model. }
    \label{fig:Rg_results}
\end{figure}

The results for the normalized radius of gyration $\langle \widetilde{R}_g(n) \rangle$ as a function of subchain length $n$ for the six coarse-grained protein models and the dataset of x-ray crystal structures are shown in Fig.~\ref{fig:Rg_results} (a). To quantify differences in the radius of gyration between each model and the x-ray crystal structure dataset, we compute the normalized mean-squared error (MSE) in $R_g(n)$:
\begin{equation}
\label{mse}
    \mathrm{MSE(\langle \widetilde{R}_g \rangle)} = \frac{ \sum\limits_{n=2}^{N} \left( \Delta \langle \widetilde{R}_{g}(n)\rangle \right)^2}{ \sum\limits_{n=2}^{N} \left( \langle \widetilde{R}_{g}^{\text{x-ray}}(n) \rangle \right)^2},
\end{equation}
where $\Delta \langle \widetilde{R}_{g}(n)\rangle = \langle \widetilde{R}_{g}^{\text{model}}(n)\rangle - \langle \widetilde{R}_{g}^{\text{x-ray}}(n) \rangle$. 

As shown in Fig.~\ref{fig:Rg_results} (a), the simplest coarse-grained model (CRW) does not recapitulate $\langle R_g(n) \rangle$ for folded proteins. $\langle R_g(n)\rangle$ for the CRW model is highly curved on a $\log$-$\log$ plot (i.e., does not possess a kink) at small $n$ and is a factor of $\sim 1.5$ smaller than $\langle R_g(n)\rangle$ for the x-ray crystal structure data at large $n$. The CRW model has the largest normalized mean-squared error relative to the x-ray crystal structure data of the six models we considered, as shown in Fig.~\ref{fig:Rg_results} (b). Similarly, $S(q)$ for the CRW model is not similar to that for the x-ray crystal structures as shown in Fig. \ref{fig:Rg_results} (c).

Introducing effective bend- and dihedral-angle potentials leads to a small, but important change in $\langle R_g(n)\rangle$ for the BADA polymer model, i.e., the appearance of a kink near $n^*\sim 10$ that separates the small- and large-$n$ regions. $\langle R_g(n) \rangle \sim n^{\nu_{1,2}}$, where $\nu_1 \sim 0.7$ for $n \lesssim n^*$ and $\nu_2 \sim 0.2$ for $n \gtrsim n^*$, which is similar to the results for the x-ray crystal structure data. However, $n^*$ for the BADA polymer model is smaller than that for the x-ray crystal structure data, and the normalized MSE in $R(n)$ for the BADA model is still quite large ($\sim 0.15$). The large MSE is caused by the fact that the persistence length of subchains in the BADA model is shorter than that for the x-ray crystal structures, and the effective bend- and dihedral-angle constraints are not sufficient to keep the subchains from over-collapsing at small $n$. 

When the amino acids are coarse-grained to include {\it single} spherical beads for both the backbone and side chain degrees of freedom (i.e., the FJSC and In Seq models), the backbone can no longer collapse as densely as found for the CRW and BADA models. For the FJSC and In Seq models, the kink location increases to $n^* \sim 30$ and their $\langle R_g(n) \rangle$ are similar to that for the x-ray crystal structure data (Fig. \ref{fig:Rg_results} (a)). The normalized MSE is $\lesssim 0.02$ for both the FJSC and In Seq polymer models (Fig. \ref{fig:Rg_results} (b)). In addition, $S(q)$ for the FJSC and In Seq models match the x-ray crystal structures much better than $S(q)$ for the BADA and CRW models (Fig. \ref{fig:Rg_results} (c)). Thus, coarse-grained protein models require at least a single side-chain bead with backbone bend- and dihedral-angle restraints to recapitulate $\langle R_g(n)\rangle$ and $S(q)$ of folded proteins. However, can the FJSC and In Seq models capture the core packing properties of proteins, such as $\langle \phi \rangle$ and $f_{\rm core}$?

Protein cores are dense packings of amino acids in the solvent-inaccessible interior of proteins, whose size and structure have been directly correlated with the stability of the protein~\cite{dill1990dominant,liang2001proteins}. Previous studies have shown that the average core packing fraction in x-ray crystal structures of proteins is $\langle \phi \rangle \approx 0.55$~\cite{gaines2016random,gaines2017packing, grigas_core_2022, grigas_protein_2025}. To identify core amino acids, we implement the software {\tt FreeSASA}~\cite{mitternacht2016freesasa} to compute the relative solvent accessible surface area (rSASA) using the Lee-Richards algorithm~\cite{lee1971interpretation}. This method employs a probe sphere to represent a solvent molecule of diameter $\widetilde{\sigma}_{\mathrm{probe}}$ that rolls over the folded protein to determine how much surface area of each amino acid it can make contact with relative to the total surface area of the fully solvated amino acid. In this work, we consider an amino acid to be in the core if $\text{rSASA} \leq 10^{-3}$, which has been previously used as an effective rSASA cutoff for identifying core amino acids~\cite{gaines2017packing,gaines_comparing_2018,treado2019void,grigas_using_2020,grigas_core_2022,grigas2024connecting, grigas_protein_2025}. Smaller diameter probes can access amino acids that are buried deeper in the protein because they can fit through smaller void spaces. Thus, we expect that as the probe shrinks, the number of amino acids found in the core will decrease and when $\widetilde{\sigma}_{\mathrm{probe}} \rightarrow 0$ the entire protein will be labeled as ``surface'', with $\langle f_{\mathrm{core}} \rangle=0$. Because proteins typically reside in water, we used a probe sphere with a diameter given by the size of a water molecule, $\sigma_{\mathrm{H_2 O}} \approx 0.73 \sigma_{\mathrm{bb}}$. Core amino acids in x-ray crystal structures are often not all nearest neighbors and instead occur in separate clusters. Motivated by this, we calculate the average {\it local} packing fraction $\langle \phi \rangle$ for each coarse-grained model conformation or x-ray crystal structure. To calculate $\langle \phi \rangle$, we perform a Voronoi tessellation and find the ratio of the volume $V_{i}$ of amino acid $i$ to the local Voronoi cell volume  $V^{\mathrm{voro}}_i$ of amino acid $i$, averaged over all $N_{\mathrm{core}}$ core amino acids: 
\begin{equation}
    \langle \phi \rangle = \frac{1}{N_{\mathrm{core}}} \sum\limits_{i=1}^{N_{\mathrm{core}}} \frac{V_{i}}{V^{\mathrm{voro}}_{i}}.
    \label{eq:avg_phi}
\end{equation}
The fraction of core amino acids is given by $f_{\rm core}=N_{\rm core}/N$. $\langle \phi \rangle$ and $\langle f_{\rm core} \rangle$ for the x-ray crystal structures are found using Voronoi tessellation, as described above, but with atomic radii used in previous studies~\cite{gaines2016random,gaines2017packing,treado2019void}. Appendix D includes additional details concerning calculations of the local packing fraction, fraction of core amino acids, and rSASA.

\begin{figure}[htbp]
    \centering
    \mbox{}%
    \adjustbox{valign=B}{\subfigure{(a)}}
    \adjincludegraphics[valign=T,width=0.87\linewidth]{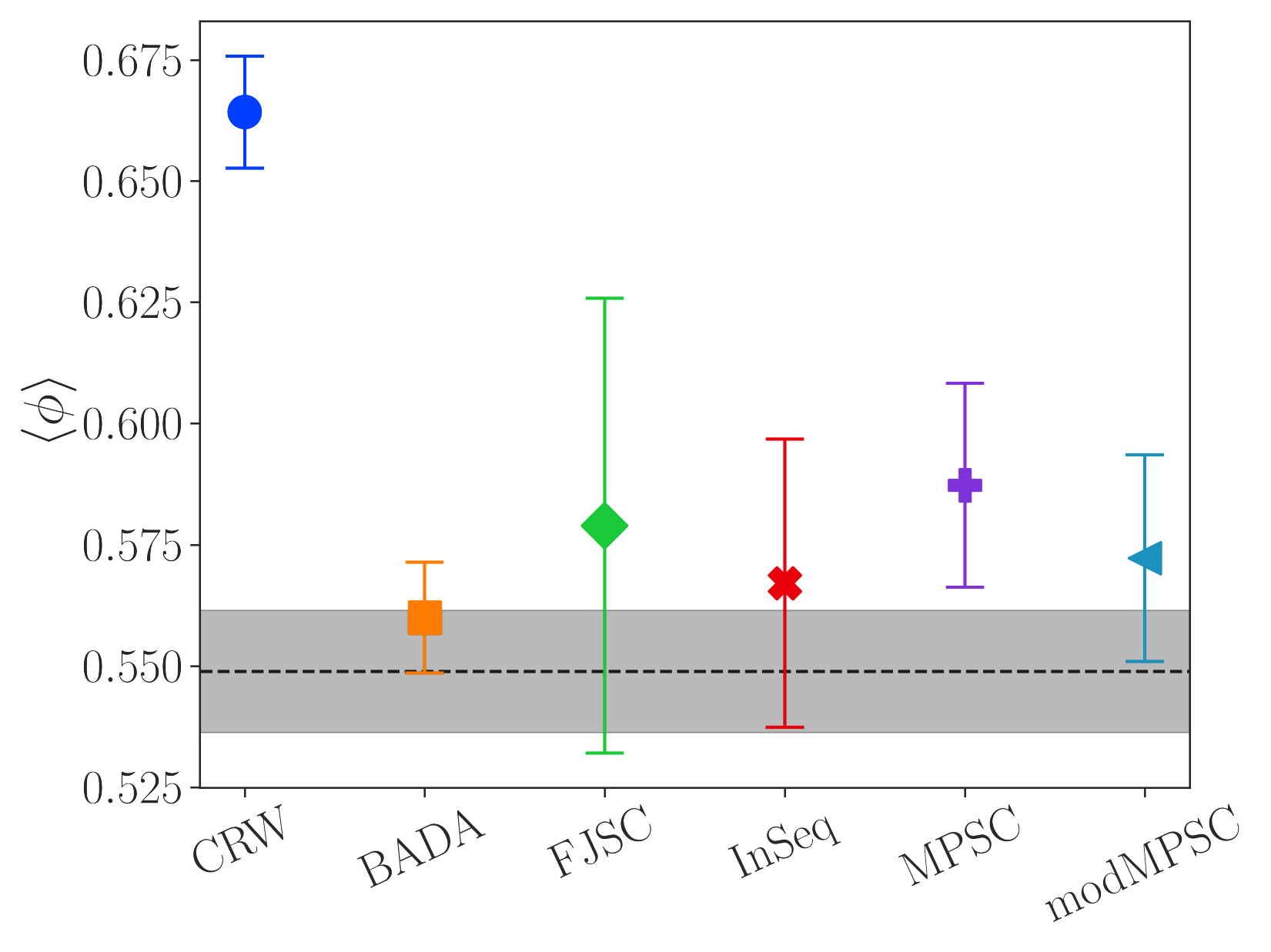}
    \par\medskip
    \adjustbox{valign=B}{\subfigure{(b)}}
    \adjincludegraphics[valign=T,width=0.87\linewidth]{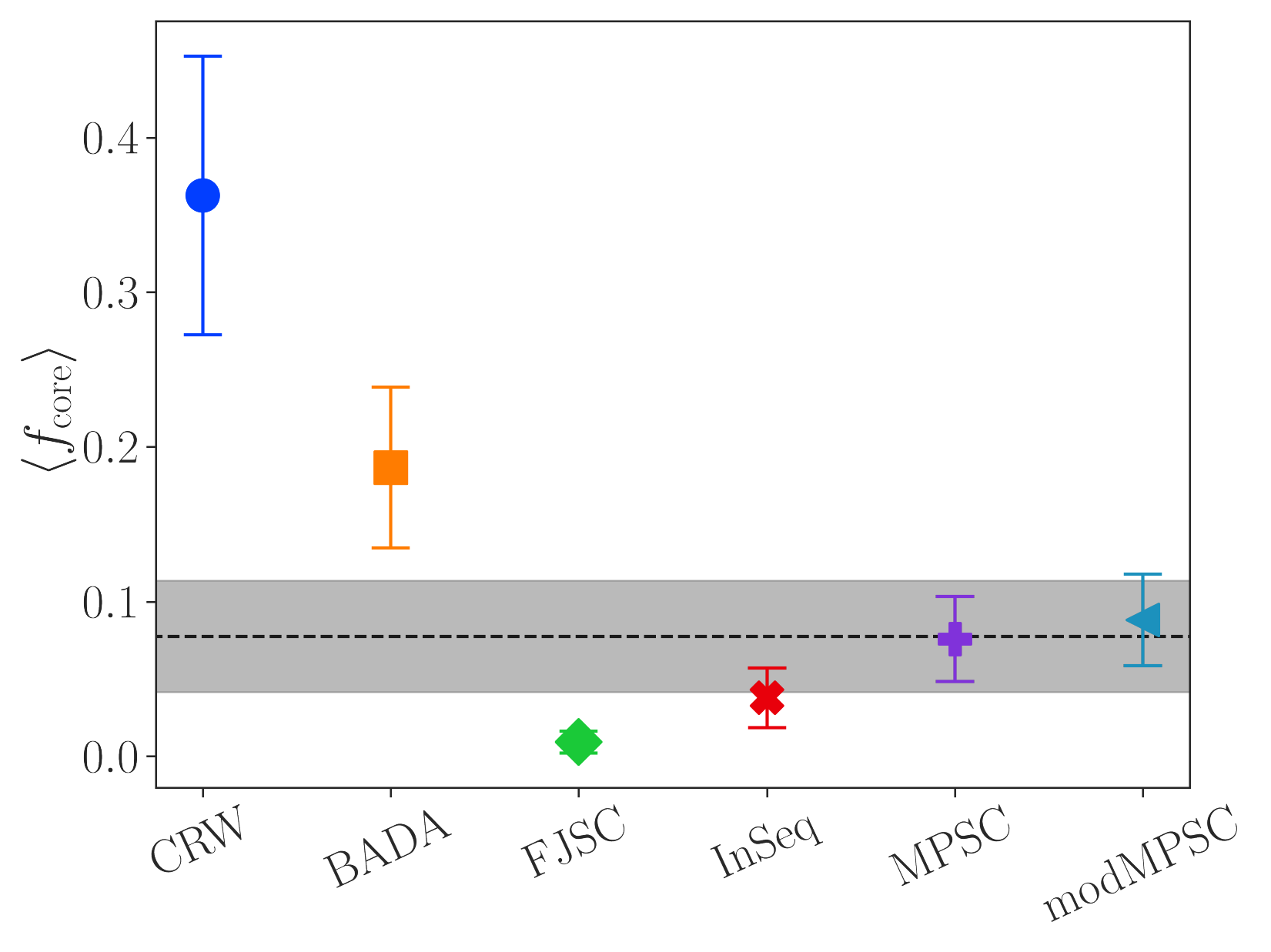}
    \caption{For each coarse-grained protein model indicated by the colors and line types in the legend, we compare (a) the average local packing fraction $\langle \phi \rangle$, and (b) the average fraction of amino acids in the core $\langle f_{\rm core} \rangle$. In panels (a) and (b), the horizontal dashed black line marks the average values for the x-ray crystal structures with $\pm 1$ standard deviation shaded in gray. The error bars for the data points in (a) and (b) represent the standard deviation of the distributions for each model.}
    \label{fig:phi_fcore_SF}
\end{figure}

\begin{figure}[htbp]
    \centering
    \mbox{}%
    \adjustbox{valign=B}{\subfigure{(a)}}
    \adjincludegraphics[valign=T,width=0.9\columnwidth]{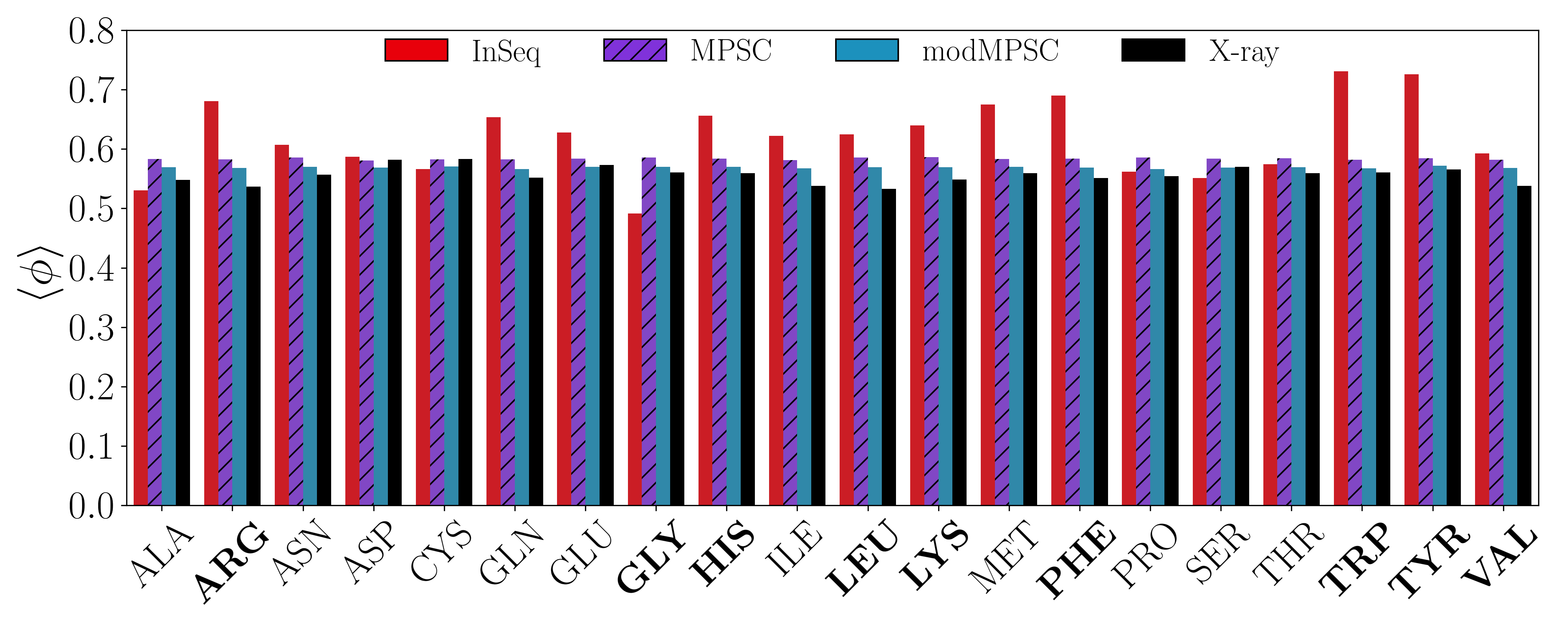}
    \par\medskip
    \adjustbox{valign=B}{\subfigure{(b)}}
    \adjincludegraphics[valign=T,width=0.9\columnwidth]{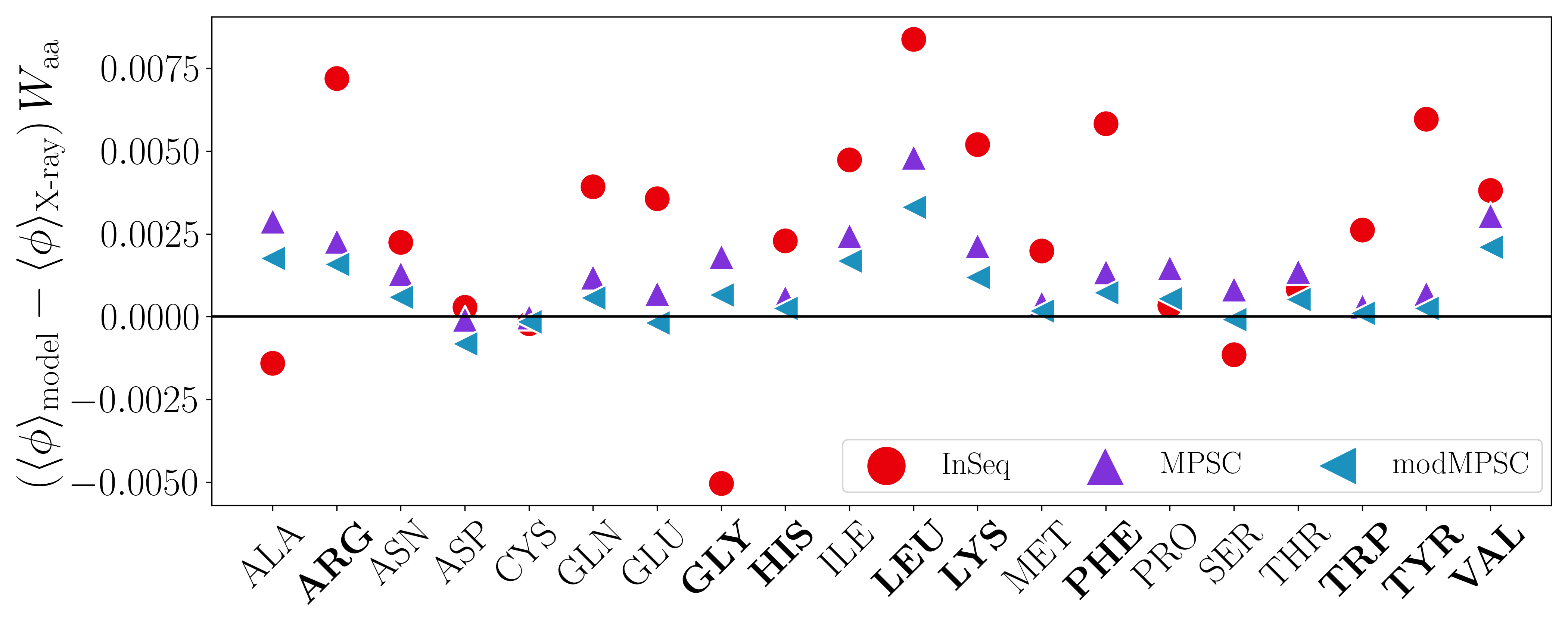}
    \caption{(a) The average packing fraction for each amino acid using the InSeq, MPSC, and modMPSC models. (b) The difference in average packing fraction between the models and the x-ray crystal structures for each protein, weighted by the abundance of each amino acid type in the dataset. As the coarse-grained models become more detailed, the packing fraction approaches the values for the x-ray crystal structures. The amino acids with more than one side chain atom in the MPSC and modMPSC models are labeled in bold.}
    \label{SI_fig:packing_fraction_by_residue}
\end{figure}

We find that the core packing properties of the FJSC and In Seq models do not strongly agree with those for x-ray structures of proteins. $\langle \phi \rangle \approx 0.57$-$0.58$ for the FJSC and In Seq models, which is similar to $\langle \phi \rangle \approx 0.55$ for the x-ray crystal structures (Fig. \ref{fig:phi_fcore_SF} (a)). However, the FJSC and In Seq models are not able to match the average packing fraction for each amino acid individually in x-ray crystal structures as shown in Fig \ref{SI_fig:packing_fraction_by_residue}(a) and (b). Further, $\langle f_{\rm core} \rangle \approx 0.02$ and $0.05$ for the FJSC and In Seq models, respectively, which indicates that the cores for the FJSC and In Seq models are significantly smaller than the cores with $\langle f_{\rm core} \rangle \approx 0.09$ for x-ray crystal structures of proteins (Fig. \ref{fig:phi_fcore_SF} (b)). Thus, we also included simulations for the MPSC and modMSPC models to show that adding multiple side chain beads can improve the coarse-grained description of the core packing properties.

We next investigate whether the slightly higher packing fraction for the MPSC and modMPSC models is the result of poorly modeling specific amino acids. In Fig.~\ref{SI_fig:packing_fraction_by_residue}, we show the average packing fraction for each amino acid type for the In Seq, MPSC, and modMPSC models compared to the x-ray crystal structures. From Fig.~\ref{SI_fig:packing_fraction_by_residue}(a), we observe that as the side chain representations become more complex, the average packing fraction for each amino acid begins to converge to the values for the x-ray crystal structures. The packing fraction for arginine, glycine, histidine, leucine, lysine, phenylalanine, tryptophan, tyrosine, and valine are all notably higher in the InSeq model than the x-ray crystal structure data. Relative to InSeq, the side‑chain geometries in the MPSC and modMPSC models yield better agreement with the amino acid packing fractions found in the x‑ray crystal structures. The reason for the remaining packing fraction error for the MPSC and modMPSC models can be seen in Fig.~\ref{SI_fig:packing_fraction_by_residue}(b), which shows the difference between the average packing fraction by residue for the models and the x-ray crystal structures weighted by the relative abundance of each amino acid in the dataset. The amino acids for the In Seq model have average packing fractions that are both greater than and less than the values for the x-ray crystal structures, while the models with more detailed side chain representations have average packing fractions that are closer to (but slightly larger than) the x-ray crystal structures. Thus, the average packing fractions for each amino acid in the modMPSC model are greater than those for the In Seq model, even though the average values for the individual amino acids for the modMPSC model are converging to the values for the x-ray crystal structures.

In summary, we find that, as expected, $\langle R_g(n)\rangle$ and $S(q)$ for the MPSC and modMPSC models are both similar to those for the x-ray crystal structures. (See Figs.~\ref{fig:Rg_results} (a) and~\ref{fig:Rg_results} (c).) $\langle f_{\rm core} \rangle$ for the MPSC model improved relative to that for the In Seq model and falls within the range for the x-ray crystal structures ($\langle f_{\rm core} \rangle \approx 0.09$). However, the core packing fraction $\langle \phi \rangle$ for the MPSC model did not move toward the value for the x-ray crystal structures ($\langle \phi \rangle \approx 0.59$). However, $\langle \phi \rangle \approx 0.57$ improves for the modMPSC model (and the average packing fractions for each amino acid individually approach those for the x-ray crystal structures), while $\langle f_{\rm core} \rangle \approx 0.09$ remains essentially unchanged and identical to the x-ray crystal structure value. These results emphasize that the side-chain representation in the modMPSC model can recapitulate the overall structural and core packing properties in x-ray crystal structures of proteins.

\section{Conclusions and Outlook \label{sec:conclusion}}

Using a series of coarse-grained protein models with increasing complexity, we identified the minimal coarse-grained models that can recapitulate several key structural properties that define folded proteins, obtained from a large dataset of more than $2500$ high-resolution x-ray crystal structures of single-chain proteins. We show that coarse-grained models with only a single backbone spherical bead cannot capture the structural properties of folded proteins. Coarse-grained protein models with a single side-chain bead (plus a single backbone bead) can recapitulate $\langle R_g(n)\rangle$ and $S(q)$, but are not able to accurately describe the core packing properties of folded proteins. Using a new coarse-grained model (modMPSC) with multiple side-chain beads, we obtain $\langle \phi \rangle$ and $f_{\rm core}$ (as well as $\langle R_g(n)\rangle$ and $S(q)$) to $< 4\%$ of the values for the x-ray crystal structures of proteins.

An important goal of this work was to identify the minimal coarse-grained protein model that can capture important structural properties of folded proteins, including the scaling of the subchain raidus of gyration $R_g(n)$, the structure factor $S(q)$, and the core packing fraction $\phi \approx 0.55$ and fraction of core amino acids $f_{\rm core} \approx 0.09$. Our results show that a purely repulsive bead‑spring backbone, plus stereochemical constraints, and a minimal side‑chain representation can be collapsed into compact structures whose ensemble‑averaged properties are statistically similar to those of x-ray crystal structures of proteins. 

We now compare the root-mean-square deviations (RMSD) of the C$_{\alpha}$ positions between the modMPSC models for each protein and the corresponding x-ray crystal structures. In the limit of both weak radial and damping forces, we have shown that the model proteins can find densely packed conformations that are similar to the corresponding x‐ray crystal structures. To estimate the $\mathrm{RMSD}$ in this limit, we placed the modMPSC model beads in positions that approximate the atomic positions in the corresponding x‐ray crystal structures and then minimized the total energy of the protein in the presence of the radial force. We find that the resulting average C$_\alpha$ $\mathrm{RMSD}$ of the core amino acids in the modMPSC models is $\sim3\,$\AA.

In previous studies, we have shown that it is possible to achieve $\lesssim 1$~\AA~core $\mathrm{RMSD}$ for all‐atom models with bend angle, backbone and side chain dihedral angle restraints using radial compressive and damping forces~\cite{grigas_protein_2025}. In future studies, we will develop {\it coarse-grained} protein models that can also achieve core $\mathrm{RMSD}$ $\lesssim$$1$~\AA. One method to lower the core RMSD is to add improper dihedral angle restraints to the side chains in the modMPSC model. Without dihedral angle restraints, the coarse-grained side chains can potentially take on non-physical conformations.  In the current modMPSC model, we increased the geometric complexity of the side chain representations for only leucine and valine. In future studies, we can increase the number of spherical beads to represent the side chains (with corresponding dihedral angle restraints) of the other amino acids. We will also determine the optimal ratio of the radial compressive and damping forces that yields core C$_{\alpha}$ $\mathrm{RMSD}$ $\lesssim 1$\AA, as well as develop short-range attractive interactions for hydrophobic amino acids in the modMPSC models that can achieve protein-specific folded states upon decreases in temperature.

Thus, we can potentially use the modMPSC model to fold all protein sequences in the human proteome. These studies would complement existing \textit{de novo} protein structure prediction methods, such as AlphaFold3~\cite{abramson_accurate_2024}. For amino acid sequences in the human proteome without experimentally determined structures, we can compare and contrast the results for folding simulations of the modMPSC model to those for AlphaFold.

\appendix
\renewcommand{\thesection}{\Alph{section}} 
\renewcommand{\thesubsection}{\thesection.\arabic{subsection}}

\section{X-ray crystal structure dataset \label{app:a}}

\begin{figure}[htbp]
    \centering
    \mbox{}%
    \adjustbox{valign=B}{\subfigure{(a)}}
    \adjincludegraphics[valign=T,width=0.9\linewidth]{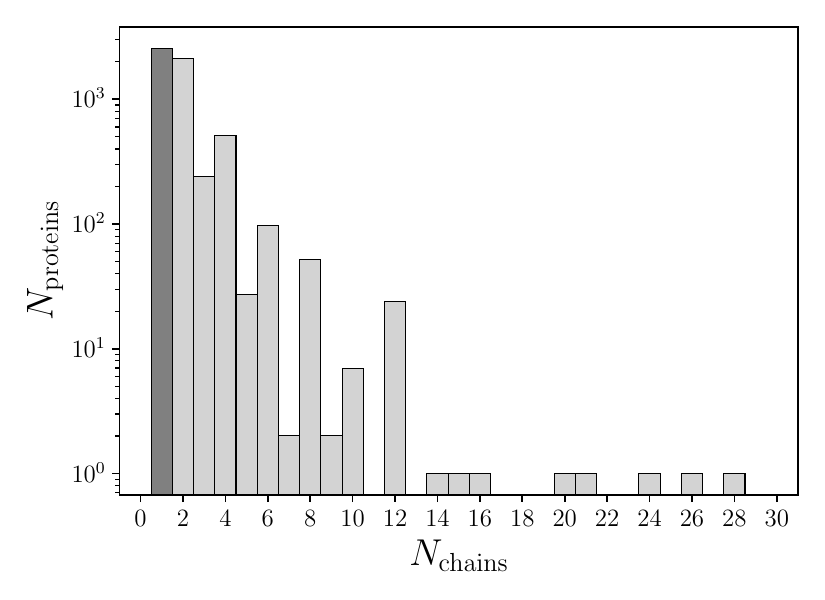}
    \par\medskip
    \adjustbox{valign=B}{\subfigure{(b)}}
    \adjincludegraphics[valign=T,width=0.9\linewidth]{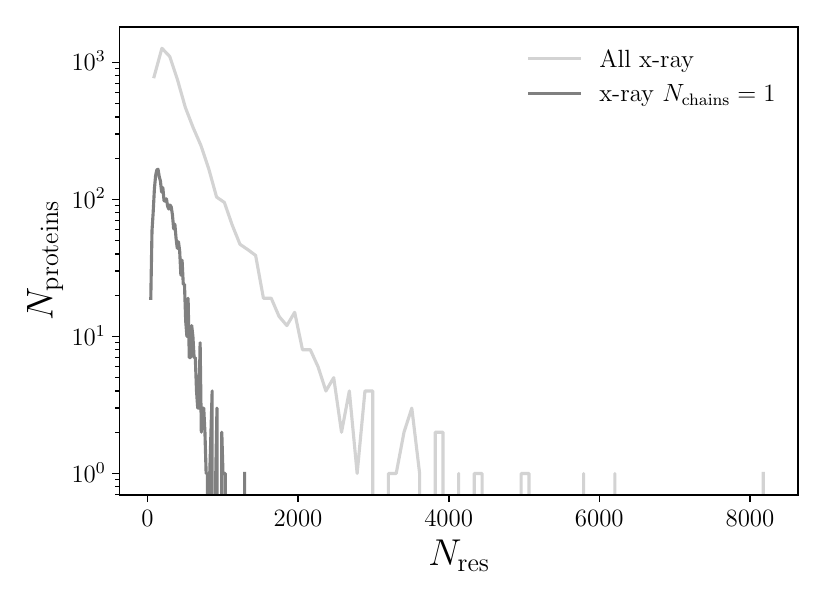}
    \caption{(a) The frequency distribution $N_{\rm proteins}$ of the number of chains $N_{\rm chains}$ in each x-ray crystal structure. The dark gray bar for $N_{\text{chains}}=1$ represents about half of the entries in the x-ray crystal structure dataset. (b) Frequency distributions $N_{\rm proteins}$ of the number of residues $N_{\rm res}$ in each x-ray crystal structure in the full dataset (black line) and the subset with $N_{\text{chains}}=1$ (dark gray line).}
    \label{SI_fig:Nchains_distr}
\end{figure}

In this work, we compare the results from the coarse-grained protein models to a subset of the Dunbrack 1.8 PISCES Protein Database of high-resolution x-ray crystal structures~\cite{wang2003pisces,wang2005pisces}. The complete dataset consists of more than $5000$ proteins ranging in length from fewer than $100$ residues to greater than $8000$ residues, with less than $50\%$ sequence similarity between structures and resolution $\leq 1.8$~\AA. Hydrogen atoms have been added to each protein x-ray crystal structure using the Reduce software~\cite{word1999asparagine}.

We cull the Dunbrack 1.8 dataset based on two criteria. First, we exclude any proteins that have unknown residues or noncanonical amino acids, such as selenocysteine (Sec)~\cite{bock1991selenocysteine,longtin2004forgotten,serrao2020selenocysteine}. For the second criterion, we only keep proteins that consist of one chain (i.e., monomeric proteins). As shown in Fig.~\ref{SI_fig:Nchains_distr} (a), single-chain proteins make up nearly half of the full Dunbrack 1.8 dataset (with $2531$ single-chain proteins). Fig.~\ref{SI_fig:Nchains_distr} (b) shows the frequency distribution of the number of residues $N_{\mathrm{res}}$ in each x-ray crystal structure for the full dataset in black and for the single-chain proteins in pink.

Among the excluded proteins are those that are composed of only a single secondary structure, e.g., only $\alpha$-helices or $\beta$-sheets. The proteins included in the analyses in the main text have an average $\langle R_g(n)\rangle$ that is similar to that for most entries in the full dataset. Thus, the subset of proteins is an accurate representation of the full dataset.

\section{Generating initial conformations \label{app:b}}

For each x-ray crystal structure and coarse-grained protein model, the initial conformation is constructed one residue at a time. We generate an initial conformation that minimizes the bond, bend-angle, and dihedral-angle potential energies, as well as the non-bonded potential energy. We begin with two backbone atoms in contact and, if the model includes explicit side chains, randomly choose a position around the respective backbone atom to attach a sidechain bead with a diameter sampled from the appropriate distribution in Fig.~4 in the main text. For the FJSC model, the diameter is chosen from a normalized distribution of all sidechain sizes regardless of the amino acid, and for the In Seq model, the diameter is chosen from the size distribution for the correct amino acid type. The sidechain bead sizes are obtained from the x-ray crystal structures in the full Dunbrack 1.8 dataset. The sidechain bead diameter is defined as the largest center-to-center distance between any two atoms plus the average of their atomic radii. We use the atomic radii from previous studies of core packing in folded proteins~\cite{gaines2016random,gaines2017packing,treado2019void}, and are listed in Table~\ref{SI_table:atom_vdW_radii}.

\begin{table}[h!]
\centering
\caption{Atomic radii used to describe the all-atom x-ray crystal structures. HX denotes a hydrogen atom bound to a carbon atom; H denotes a hydrogen atom bound to any other atom.}
\label{SI_table:atom_vdW_radii}
\begin{tabular}{lc}
\hline\hline
Atom type & Radius (\AA) \\
\hline
C  & 1.50 \\
CO & 1.30 \\
N  & 1.30 \\
O  & 1.40 \\
H  & 1.00 \\
S  & 1.75 \\
HX & 1.10 \\
\hline\hline
\end{tabular}
\end{table}

We check for overlaps each time a new bead is added to the model, but, importantly, so that we do not bias the sampling of sidechain diameters to smaller values, we allow for small overlaps between sidechain beads when initially building the model conformation. Starting with the third backbone bead, a new backbone position is chosen randomly in the subspace that minimizes the bond length and bend angle potential energies, without overlapping another backbone bead. Depending on the model, subsequent backbone beads are also selected so that the dihedral angles sample $\mathcal{P}(\psi_{ijkl}) \propto \mathrm{e}^{-\widetilde{U}_{\mathrm{dh}}(\psi_{ijkl})}$. After all beads have been added for a given coarse-grained model and x-ray crystal structure, damped molecular dynamics simulations are carried out to remove bead overlaps. 

The MPSC model is initialized starting with conformations from the In Seq model. The side chain representations for the In Seq model are modified to have $0$-$5$ spherical beads as used for the Martini3 model \cite{souza_martini_2021}. Compared to the In Seq model, the MPSC model changes the side chain representations for glycine, histidine, arginine, lysine, phenylalanine, tryptophan, and tyrosine. The spherical beads for the MPSC side chain representations of each amino acid are the same size and the spherical beads are scaled so that the sum of their diameters matches the diameter of the single side chain bead for the In Seq model. The modMPSC model is initialized in the same manner as the MPSC model. However, the modMPSC model modifies the side chain representations for valine and leucine to include two spherical beads instead of one. After changing the side chains for the MPSC and modMPSC models, we run a short NVE simulation for each of the coarse-grained proteins to ensure a novel conformation before the production simulation.

\section{Dihedral angle potential energy \label{app:c}}

The dihedral angle potential energy in Fig.~3 (b) in the main text was constructed from the probability distribution $\mathcal{P}(\psi_{ijkl})$ of dihedral angles between four consecutive $C_{\alpha}$ atoms using Langevin dynamics of the united atom (UA) model for proteins in previous work~\cite{smith2014calibrated}. The dihedral angle potential energy is obtained from $U_{\text{dh}} \propto -k_bT \left\langle \ln \mathcal{P}(\psi_{ijkl}) \right\rangle$ and fit with a fourth-order Fourier series, with the coefficients listed in Table~\ref{SI_table:dih_pot_coeff}. The peak near $\psi_{ijkl}=-60^\circ$ and the plateau in the range $0^\circ<\psi_{ijkl}<120^\circ$ can be attributed to secondary structure in proteins. 

\begin{equation}
    U_{\mathrm{dh}}(\psi_{ijkl}) = U_{\mathrm{da}} \sum\limits_{\langle ijkl\rangle} \sum\limits_{s=1}^{4} \Bigl[ A_s \cos\left( s \, \psi_{ijkl}\right) + B_s \sin\left( s \, \psi_{ijkl}\right)  \Bigr].
\label{SI_eq:dihedral_PE}
\end{equation}

\begin{table}[h!]
\centering
\begin{tabular}{ccccc}
\hline \hline
$s$ & 1 & 2 & 3 & 4 \\ \hline
$A_s$ ($\times \, 5\cdot10^{-7}$) & 70.5 & -31.3 & -7.9 & 4.1 \\ 
$B_s$ ($\times \, 5\cdot10^{-7}$) & -17.5 & -9.3 & 3.0 & 3.0 \\ \hline \hline
\end{tabular}
\caption{Coefficients of the backbone dihedral angle potential energy $\boldsymbol{U_{\mathrm{dh}}}$.}
\label{SI_table:dih_pot_coeff}
\end{table}

All physical quantities, including the dihedral angle potential energy, are made dimensionless using the energy, mass and length units: $\epsilon_{\mathrm{rep}}$, $m$, and $\sigma_{\mathrm{bb}}$. The values used in the simulations are listed in Table~\ref{SI_table:simulation_parameters}.

\begin{table}[h]
\centering
\begin{tabular}{cc}
\hline \hline
Parameter & Value \\ \hline
$\widetilde{U}_{\mathrm{bb}}$  & 1.0 \\ 
$\widetilde{U}_{\mathrm{ba}}$  & 1.0 \\ 
$\widetilde{U}_{\mathrm{da}}$  & 1.0 \\ 
$\widetilde{F}_{\mathrm{cent}}$  & $10^{-4}$ \\ \hline \hline
\end{tabular}
\caption{Dimensionless simulation parameters.}
\label{SI_table:simulation_parameters}
\end{table}

\section{Calculating core properties \label{app:d}}

\subsection*{Fraction core, $f_{\text{core}}$}
\begin{figure}[htbp]
    \centering
    \mbox{}%
    \adjustbox{valign=B}{\subfigure{(a)}}
    \adjincludegraphics[valign=T,width=0.9\linewidth]{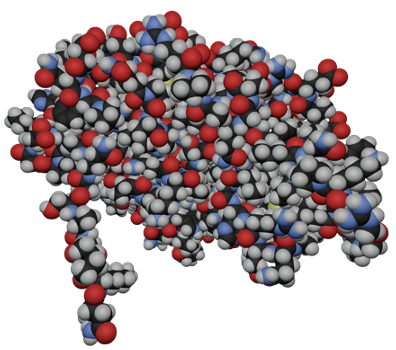}
    \par\medskip
    \adjustbox{valign=B}{\subfigure{(b)}}
    \adjincludegraphics[valign=T,width=0.9\linewidth]{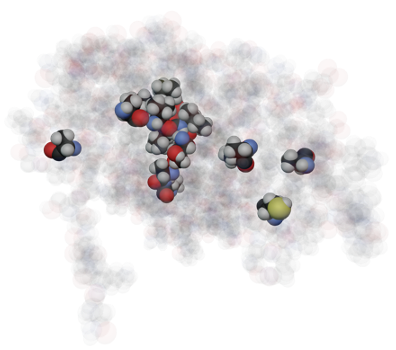}
    \caption{(a) All-atom representation of PDBID: 5TKW. (b) The core of protein 5TKW is highlighted by making all non-core atoms transparent. This protein has a core that is not contiguous, which is common for folded proteins.}
    \label{SI_fig:all_atom_and_core_examples}
\end{figure}

The first step in calculating the fraction of core amino acids ($f_{\rm core}$) is to identify which residues are in the core. In this work, we define a residue to be in the protein core if the relative solvent accessible surface area (rSASA) $\text{rSASA} \leq 10^{-3}$. A probe particle of diameter $\sigma_{\text{probe}}$ ($\sigma_{\text{probe}} = 0.73\sigma_{\text{bb}}$ in this work) represents the solvent and moves on the surface and through the geometrically accessible void space of the protein. rSASA is the ratio of the surface area of the residue that the probe can make contact with to the surface area of the full residue, removed from the protein and fully solvated. An example of the all-atom structure for 5TKW is shown in Fig. \ref{SI_fig:all_atom_and_core_examples}(a), with its identified core shown in Fig \ref{SI_fig:all_atom_and_core_examples}(b). Once the core residues are identified, $f_{\text{core}}$ is calculated by dividing the number of core amino acids by the total number of amino acids in the protein.

\subsection*{Packing fraction, $\phi$}
Calculating the packing fraction ($\phi$) starts with identifying the core residues using the same method as that used for $f_{\text{core}}$. In addition to rSASA, we check for surface residues using a radical Voronoi tessellation. In a Voronoi tessellation, if a point lies on the collapsed polymer surface, the volume of its cell will depend on the bounding box. Any Voronoi cell volume that remains constant after scaling the boundary is labeled as being in the interior of the polymer. The complement of the union of surface beads from the rSASA and Voronoi methods defines the core residues (and their Voronoi cells) used for calculating the packing fraction. Any residues identified in the core are then used to compute the packing fraction of the core.

For all collapsed models and x-ray crystal structures, we treat each sphere indiviudally, whether it represents a side chain atom or backbone atom. If a pair of atoms has negligible overlap, each Voronoi cell will contain the entire sphere. If there are overlaps between two spheres, the Voronoi plane will intersect the spheres. The collapsed models have negligible overlaps, but the all-atom x-ray crystal structures possess both inter-residue and intra-residue atomic overlaps. There are several ways to compute the net volume of spheres in a residue while accounting for overlaps, such as the inclusion-exclusion principle, which requires adding and subtracting the overlap volumes of pairs, triples, etc.. We chose to use a Monte Carlo method for calculating volumes of amino acids. We place the core residues in a box and randomly sample points in the box. In the case of a point that falls in the overlap region between a core atom and a surface atom, we check on which side of the Voronoi plane the point falls. If the point is on the core side of the Voronoi plane, it is counted as in the core, otherwise it is outside of the core. Given $N$ points sampled in the bounding box of volume $V_{\text{box}}$, if $N_{\text{in}}$ are found to fall in the amino acid on the core side of any atoms near the core-surface boundary, the volume of the amino acid is  
\begin{equation*}
    V_{\text{res}} \approx \frac{N_{\text{in}}}{N} V_{\text{box}}.
\end{equation*}
After calculating $V_{\text{res}}$ for all core residues in the protein, we can determine the packing fraction $\phi$ by averaging the local packing fractions of each core residue,
\begin{equation*}
    \phi_{\text{res}} \approx \frac{V_{\text{res}}}{V^{\text{voro}}_{\text{res}}},
\end{equation*}
and $\phi = \langle \phi_{\text{res}} \rangle$.
As the number of test points grows, the volume computation can be made as accurate as desired, and the packing fraction converges. In addition to using a high density of sampled points (20,000 $\text{points}/\sigma^3_{\text{bb}}$), we also compute the final local packing fraction for each residue by averaging over $50$ independent realizations.
\par\vspace{0pt plus -1fil}

\bibliography{main}

\end{document}